\documentclass[10pt,journal,compsoc]{IEEEtran}
\IEEEoverridecommandlockouts

\usepackage{multirow}

\usepackage{cite}
\usepackage{amsmath,amssymb,amsfonts}
\usepackage{algorithmic}
\usepackage{graphicx}
\usepackage{textcomp}

\usepackage{xspace}
\usepackage[table]{xcolor}

\usepackage{listings}
\usepackage{color}
\usepackage{todonotes}

\usepackage{comment}
\usepackage{hyperref}
\usepackage{enumitem}

\usepackage{tikz}

\newcommand*\circled[1]{\tikz[baseline=(char.base)]{
    -                \node[shape=circle,draw,inner sep=1pt,minimum size=2pt] (char) {\scriptsize{#1}};}}

\newcommand\rev[1]{\textcolor{black}{#1}}

\definecolor{codegreen}{rgb}{0,0.6,0}
\definecolor{codegray}{rgb}{0.5,0.5,0.5}
\definecolor{codepurple}{rgb}{0.58,0,0.82}
\definecolor{backcolour}{rgb}{0.95,0.95,0.92}
\definecolor{codeblue}{rgb}{0,0.1,0.95}
 
\lstdefinestyle{mystyle}{
    linewidth=240,
    backgroundcolor=\color{backcolour},   
    commentstyle=\color{codegreen},
    keywordstyle=\color{codeblue},
    numberstyle=\tiny\color{codegray},
    stringstyle=\color{codepurple},
    basicstyle=\footnotesize\fontsize{6}{11}\ttfamily,
    breakatwhitespace=false,         
    breaklines=true,                 
    captionpos=b,                    
    keepspaces=true,                 
    numbers=left,                    
    numbersep=5pt,                  
    showspaces=false,                
    showstringspaces=false,
    showtabs=false,                  
    tabsize=2
}
\def\BibTeX{{\rm B\kern-.05em{\sc i\kern-.025em b}\kern-.08em
    T\kern-.1667em\lower.7ex\hbox{E}\kern-.125emX}}

\begin{document}

\newcommand{\cachedump}{CacheFlow\xspace}
\newcommand{\dumper}{Shutter\xspace}
\newcommand{\trigger}{Trigger\xspace}
\newcommand{\filter}{Filter\xspace}

\newcommand{\ramindex}{\texttt{RAMINDEX}\xspace}

\title{Observing the Invisible: Live Cache Inspection \\for
  High-Performance Embedded Systems}

\author{
   \IEEEauthorblockN{
     Dharmesh Tarapore\IEEEauthorrefmark{1},
     Shahin Roozkhosh\IEEEauthorrefmark{1},
     Steven Brzozowski\IEEEauthorrefmark{1} and
     Renato Mancuso\IEEEauthorrefmark{1}}\\
   \IEEEauthorblockA{\IEEEauthorrefmark{1}Boston University, USA \{dharmesh, shahin, sbrz, rmancuso\}@bu.edu}
 }



\IEEEtitleabstractindextext{%
\begin{abstract}
The vast majority of high-performance embedded systems
implement multi-level CPU cache hierarchies. But the exact behavior
of these CPU caches has historically been opaque to system designers.
Absent expensive hardware debuggers, an understanding of cache makeup
remains tenuous at best. This enduring opacity further obscures the complex
interplay among applications and OS-level components, particularly as they
compete for the allocation of cache resources.
Notwithstanding the relegation of cache comprehension to proxies such as
static cache analysis, performance counter-based profiling, and cache hierarchy
simulations, the underpinnings of cache structure and evolution continue to elude
software-centric solutions.

In this paper, we explore a novel method of studying cache contents
and their evolution via snapshotting. Our method complements extant
approaches for cache profiling to better formulate, validate, and refine
hypotheses on the behavior of modern caches. We leverage cache
introspection interfaces provided by vendors to perform live cache
inspections without the need for external hardware.
We present \cachedump, a proof-of-concept Linux kernel module which
snapshots cache contents on an NVIDIA Tegra TX1 SoC (system on chip).
\end{abstract}
\begin{IEEEkeywords}
cache, cache snapshotting, ramindex, cacheflow, cache debugging
\end{IEEEkeywords}}

"This work has been submitted to the IEEE for possible publication.
Copyright may be transferred without notice, after which this version
may no longer be accessible."

\maketitle

\section{Introduction}

The burgeoning demand for high-performance embedded systems
among a diverse range of applications such as telemetry, embedded
machine vision, and vector processing has outstripped the capabilities
of traditional micro-controllers. For manufacturers, this has engendered a
discernible shift to system-on-chip modules (SoCs). Coupled with their
extensibility and improved mean time between failures,
SoCs offer improved reliability and functionality.
To bridge the gap between increasingly faster CPU speeds and
comparatively slower main memory technologies (e.g. DRAM) most
SoCs feature cache-based architectures. Indeed, caches allow modern
embedded CPUs to meet the performance requirements of
emerging data-intensive workload. At the same time, the strong need for
predictability in embedded applications has rendered analyses of caches
and their contents vital for system design, validation, and certification. 

Unfortunately, this interest runs counter to the general desire to
abstract complexity. Cache mechanisms and policies are ensconced
entirely in hardware, to eliminate software interference and encourage
portability. Consequently, software-based techniques used to study
caches suffer from several shortcomings, which we detail in Section
\ref{sec:comparison_hw}.  

  
  

In contrast, we propose \cachedump: a technique that can be implemented
in software and in existing high-performance SoCs to extract and analyze
the contents of cache memories. \cachedump can be deployed in a live
system without the need for an external hardware debugger. By periodically
sampling the cache state, we show that we can reconstruct the behavior of multiple
applications in the system; observe the impact of scheduling policies; and study
how multi-core contention affects the composition of cached content.

Importantly, our technique is not meant to replace other cache
analysis approaches.  Rather, it seeks to supplement them with
insights on the exact behavior of applications and system components
that were not previously possible.  While in this work we specifically
focus on last-level (shared) cache analysis, the same technique can be
used to profile private cache levels, TLBs, and the internal states of
coherence controllers. In summary, we make the following
contributions:

\begin{enumerate}[leftmargin=*]
\item This is the first paper to describe in detail an interface,
  namely \ramindex, available on modern embedded CPUs that can be used
  to inspect the content of CPU caches. Despite its usefulness, the
  interface has received little to no attention in the research
  community thus far;
  
\item We present a technique called \cachedump to perform cache
  content analysis via event-driven snapshotting;
  
\item We demonstrate that the technique can be implemented on modern
  hardware by leveraging the \ramindex interface and propose a
  proof-of-concept open-source Linux implementation\footnote{Our
    implementation can be found at: \textit{\url{https://github.com/weirdindiankid/cacheflow}}}

\item We describe how to correlate information retrieved via cache
  snapshotting to user-level and kernel software components deployed
  on the system under analysis;
  
\item We evaluate some of the insights provided by the proposed
  technique using real and synthetic benchmarks.
  
\end{enumerate}

The rest of this paper is organized as follows: in Section II, we document
related research that inspired and informed this paper. In Sections III and IV,
we provide a bird's-eye view of the concepts necessary to understand the
mechanics of \cachedump. In Sections V and VI, we detail the fundamentals of
\cachedump and its implementation. Section VII outlines the experiments we
performed and documents their results. We also examine the implications of 
those results. Section VIII concludes the paper with an outlook on future 
research.

\section{Related Work}

\par Caches have a significant impact on the temporal behavior of
embedded applications. But their design\textemdash oriented toward programming
transparency and average-case optimization\textemdash makes performance impact
analysis difficult. A plethora of
techniques have approached cache analysis from multiple angles. We
hereby provide a brief overview of the research in this space.

{\bf Static Cache Analysis} derives bounds on the access time of 
memory operations when caches are present
\cite{survey_sttc_anlsys_2016, wcet_survey_mtd_tools_2008,
  grund2012static}. Works in this domain study the set of possible
cache states in the control-flow graph (CFG) of applications. Abstract
interpretation is widely employed for static cache analysis, as first
proposed in \cite{ferdinand1999efficient} and
\cite{cousot2004basic}. For static analysis to be carried out, a
precise model of the cache behavior is required. Techniques that
consider Least-Recently Used (LRU), Pseudo-LRU, and
FIFO replacement policies
~\cite{white1997timing, reineke2007timing, grund2012static,
  guan2013fifo, grund2010precise} have been studied. 
 





\par{\bf Symbolic Execution} is a software technique for feasible path
exploration and WCET analysis~\cite{chu2011symbolic, chu2016precise}
of a program subject to variable input vectors. It proffers a middle ground
between simulation and static analysis. An interpreter follows the program;
if the execution path depends on an unknown, a new symbolic executor is
forked. Each symbolic executor stands for many actual program runs whose
actual values satisfy the path condition. 





As systems grow more complex, {\bf Cache Simulation} tools are
essential to study new designs and evaluate existing ones.
Simulation of the entire processor --- including cores, cache
hierarchy, and on-chip interconnect --- was proposed
in~\cite{gem5_2011,tejas_2015,manifold_simulator_2014,
  valgrind_sim_2003}. Simulators that only focus on the cache
hierarchy were studied in~\cite{wang2013fast,
  coutinho_mscsim_simulator_2006, arda_ds3_2020}. Depending on the
component under analysis, simulations abound.
In the (i) execution-driven approach, the program to be traced
runs locally on the host platform; in the (ii) emulation-driven
approach, the target program runs on an emulated platform and
environment created by the host; finally, in the (iii) trace-driven
approach, a trace file generated by the target application is fed into
to the simulator. An excellent survey reviewing 28 CPU cache
simulators was published by Brais
et. al~\cite{brais_survey_cache_simulators_2020}. The most popular is
perhaps Cachegrind that belongs to the Valgrind Suite
\cite{valgrind_simulator_2007}.

\par{\bf Statistic Profiling} is performed by leveraging performance
monitoring units (PMUs) integrated in modern processors. PMUs
can monitor a multitude of hardware events that occur as applications
execute on the platform. Unlike the aforementioned strategies,
sampling the PMU provides information on the real behavior of
the hardware platform. As such, a number of works have used statistic
profiling to study memory-related performance
issues~\cite{levinthal2009performance, treibig2012performance,
  molka_detect_memory_hw_perfcntr_2017}. High level libraries such as
PAPI~\cite{mucci1999papi,browne2000portable},
Likwid~\cite{treibig2010likwid} and numap~\cite{selva2016numap}
provide a set of APIs to ease the use of
PMUs. 

Despite the seminal results achieved in the last decade in the cache
analysis and profiling techniques described thus far, a few important
limitations are worth noting. Techniques that rely on cache models ---
i.e. static analysis, symbolic execution, simulation --- work under
the assumption that the employed models accurately
represent the true behavior of the hardware. Unfortunately, complex
modern hardware often deviates from \emph{textbook} models in
unpredictable ways. Access to event counters
subsequently only reveals partial information on the actual state
of the cache hierarchy. 

The technique proposed in this paper is meant to complement the
analysis and profiling strategies reviewed thus far. It does so by
allowing system designers to \emph{snapshot} the actual content of CPU
caches. This, in turn, enables a new set of strategies to
extract/validate hardware models, or to conduct application and
system-level analysis on the utilization of cache resources. Unlike
works that proposed cache snapshotting by means of hardware
modifications~\cite{vishnoi2009online, buck2006new,
  panda2010enhancing} our technique can be entirely implemented in
software and leverages hardware support that already exists in a broad
line of high-performance embedded CPUs.

\section{Background}

In this section, we introduce a few fundamental concepts required to
understand \cachedump's inner workings. First,
we review the structure and functioning of multi-level set-associative
caches. Next, we briefly explore the organization of virtual memory in the
target class of SoCs.

\subsection{Multi-Level Caches}

Modern high-performance embedded processors implement multiple levels
of caching. The first level (L1) is the closest to the CPU and its
contents are usually private to the local processor. Cache misses in L1
trigger a look-up in the next cache level (L2), which can be still
private, shared among a subset (cluster) of cores, or globally shared
by all the cores. Additional levels (L3, L4, etc.) may also be present.

The last act before a look-up in the main memory is to query
the \textit{Last-Level Cache} (LLC). Without loss of generality, and to
be more in line with our implementation, we consider the typical cache
layout of ARM-based processors. That is, we assume private L1 caches
and a globally shared L2, which is also the LLC.
\par{\bf Set-associativity:} Caching at any level typically
follows a \textit{set-associative} scheme. A set-associative cache
with associativity $W$, features $W$ ways, where each way has an
identical structure. A cache with total size $C_S$ is thus structured
in $W$ ways of size $W_S=C_S/W$ each.  Caches store multiple blocks of
consecutive bytes in \textit{cache lines} (a.k.a. \textit{cache
  blocks}). We use $L_S$ to indicate the number of bytes in each cache
line. Typical line sizes are 32 or 64 bytes. $S=W_S/L_S$ denotes the
number of lines in a way or \textit{sets} in a cache.



When the CPU accesses a \textit{cacheable} memory location, the memory
address determines how the cache look-up (or allocation) is performed.
In the case of a cache 
miss, the least-significant bits of the address encode the specific
byte inside the cache line and are called \textit{offset} bits. For
instance, in systems with $L_S=64$ bytes, the offset bits are
[5:0]. The second group of bits in the memory address encodes the
specific cache set in which the memory content can be cached. Since we
have $S$ possible sets, the next $log_2S$ bits after the offset bits
select one of the possible sets and are called \textit{index}
bits. Finally, the remaining bits, called \textit{tag} bits, are
stored alongside cached content to detect cache hits after
look-up.

\par{\bf Virtual and Physical Caches:} Addresses used for cache
look-ups can be physical or virtual. In the vast majority of
embedded multi-core systems, tag bits are physical address bits
signifying a \textit{physically-tagged} cache. In shared caches
(e.g. L2), index bits are also derived from physical addresses. For
this reason, they are said to be \textit{physically-indexed},
\textit{physically-tagged} (PIPT) caches~\cite{arm-cortex-a53,
  arm-cortex-a57, arm-cortex-a72, arm-cortex-a15,
  arm-cortex-a9}.

\subsection{Memory Management \& Representation}

\par{\bf Virtual Memory} Modern computing architectures rely on
software and hardware support to map virtual addresses used by
processes to physical addresses represented in the
hardware. Giving processes distinct views of the system's
memory obviates problems stemming from fragmentation or
limited physical memory. The operating system manages the
translation between virtual and physical addresses through
a construct known as the \textit{page table}. When a process tries
to reference a virtual address, the operating system first checks
for that address' presence in the referencing process' address
space. If present, the system then checks for the page's
presence in memory via its page table entry (PTE). From there,
the address is either resolved to a physical address, or the
page is brought into main memory and then resolved.

\par{\bf Virtual Memory Areas}:
When a process terminates in Linux, the kernel is tasked with freeing
the process' memory mappings. Older versions of Linux accomplished
this by iterating through a chain of reverse pointers to page tables
that referenced each physical frame. The mounting intractability of this
approach spurred the development of \emph{virtual memory areas}, or VMAs.

VMAs are contiguous regions of virtual memory represented as a range of
start and end addresses. They simplify the kernel's management of a
process' address space, thus facilitating granular control permissions
on a per-VMA basis. VMAs record frame to page mappings on a per-VMA basis
(as opposed to on a per-frame basis) and were incorporated into the kernel,
as of version 2.6 ~\cite{rmap_kdev}.

\section{The \ramindex Interface}
\label{sec:ramindex}
The \ramindex interface was originally introduced on the ARM
Cortex-A15~\cite{arm-cortex-a15} family of CPUs and is currently
available on the high-performance line of ARM embedded
processors. These include ARM Cortex A15~\cite{arm-cortex-a15}, ARM
Cortex A57~\cite{arm-cortex-a57}, ARM Cortex
A76~\cite{arm-cortex-a76}, and the recently announced ARM Cortex
A77~\cite{arm-cortex-a77}. Table~\ref{tab:hw_support} reviews the
availability of the \ramindex interface across ARM CPUs and provides a
few notable examples of known SoCs equipped with such CPUs. Given the
consistent support for the \ramindex interface across the
high-performance line of ARM processors, there is good indication that
\ramindex will continue to be supported in future families of CPUs.


\renewcommand{\arraystretch}{1.2}

\hspace{-0.5cm}
\begin{table}
  \caption{Availability of \ramindex in ARM Cortex-A CPUs.}
  \resizebox{1\columnwidth}{!}{%
    \begin{tabular}{@{}l|l|l|l|lll}
    \textbf{CPU} &
    \textbf{Max. Freq.} &
    \textbf{Release} &
    \textbf{Privilege Level} &
    \textbf{Notable SoCs}
    \\ \hline \hline
    Cortex-A9 (32-bit) &
    800 MHz &
    2007 &
    Not supported &
    \begin{tabular}[c]{@{}l@{}}Xilinx Zynq-7000\\
      Nvidia Tegra, 2, 3, 4i\end{tabular}
      \\ \hline
      Cortex-A15 (32-bit) &
      2.5 GHz &
      2011 &
      EL1 or Higher &
      \begin{tabular}[c]{@{}l@{}}Nvidia Tegra 4, K1\\
        MediaTek MT8135/V\end{tabular}
        \\ \hline
        Cortex-A53 (64-bit)&
        1.5 GHz &
        2012 &
        Not supported &
        \begin{tabular}[c]{@{}l@{}}Intel Stratix 10\\
          Xilinx ZynqMP\end{tabular}
          \\ \hline
          Cortex-A57 (64-bit) &
          1.6 GHz &
          2012 &
          EL1 or Higher &
          \begin{tabular}[c]{@{}l@{}}Nvidia Tegra X1\\
            Nvidia Tegra X2\end{tabular}
            \\ \hline
            Cortex-A72 (62-bit) &
            2.5 GHz &
            2015 &
            EL1 or Higher &
            \begin{tabular}[c]{@{}l@{}}MediaTek Helio X2x, MT817x\\
              Xilinx Versal\end{tabular}
              \\ \hline
              Cortex-A76 (64-bit) &
              3 GHz &
              2018 &
              EL3 &
              MediaTek Helio G90
              \\ \hline
              Cortex-A77 (64-bit) &
              3.3 GHz &
              2019 &
              EL3 &
              MediaTek Dimensity 1000
              \\ \hline
    \end{tabular}
}
\label{tab:hw_support}
\vspace{-0.4cm}
\end{table}

We now make specific reference to the \ramindex interface available
on ARM Cortex-A57 CPUs and used in our experiments. These CPUs belong
to the ARMv8-A architecture and support two main modes of operations:
64-bit mode (AArch64) and 32-bit mode (AArch32). We base our
discussion on target machines operating in AArch64
mode.

\ramindex operates on a set of 10 32-bit wide system
registers which are local to each of the processing cores. These
registers are read via the \emph{move from
  system register} (\texttt{mrs}) instruction. Similarly, write
operations can be performed using the \emph{move to system register}
(\texttt{msr}) instruction.

The main interface is the eponymous \ramindex register.  In a
nutshell, the CPU writes a command to the \ramindex register by
specifying (1) the target memory to be accessed, and (2) appropriate
coordinates to access the target memory. The result of the command is
then available in the remaining 9 registers and can be read. Of these,
the first group of registers named \texttt{IL1DATA$n$\_EL1}, with $n
\in \{0, \ldots, 3\}$ holds the result of any operation that accesses
the instruction L1 cache. The second group of registers, namely
\texttt{DL1DATA$n$\_EL1} with $n \in \{0, \ldots, 4\}$, holds the
result of accesses performed to L1 or L2 cache data entries. The
resources that can be accessed through the \ramindex interface
(see~\cite{arm-cortex-a57} for exact details of the valid commands)
include (1) L1 instruction and data cache content, both tag and data
memories; (2) L1 instruction cache branch predictor and indirect jump
predictor memories\footnote{These are three different memory
  resources, i.e. the indirect predictor memory, the Branch Target
  Buffer (BTB) and the Global Hostory Buffer (GHB).}; (3) L1
instructions and data translation look-aside buffers (TLBs); (4) L2
tag and data memories; (5) L2 TLB; (6) L2 Error Correction Code (ECC)
memory; and (7) L2 snoop-control memory yielding information about the
Modified/Owned/Exclusive/Shared/Invalid (MOESI) state of each cache
line.
 
As can be noted, the coverage of the \ramindex interface is quite
extensive. In this paper we decided to focus specifically on the L2
cache which is shared among all the cores. Because we are specifically
interested in the relationship between cache state and memory owned by
applications, we program the \ramindex interface to access the L2 tag
memory. The coordinates to retrieve the content of a specific entry
are expressed in terms of (1) cache way number, and (2) cache set
number to be accessed. Because the L2 is a PIPT cache, the set number
corresponds to the index bits of a physical address. The content of
the tag memory returned on the \texttt{DL1DATA$n$\_EL1} registers
contains the remaining bits of the physical address cached at the
specified coordinates, and an indication that the entry is indeed
valid.



\subsection{Security Considerations}
Among the four privilege levels E0-E3 of ARMv8(-A), on ARM Cortex-A57
and Cortex-A72 CPUs, \ramindex is available at all exception levels,
excluding EL0, i.e. the lowest privilege level (see
Table~\ref{tab:hw_support}).  This has important implications. For
example, \ramindex could be used to expose memory and cache contents
of other guest operating systems sharing the same hardware.
For this reason, while in ARM
Cortex-A57, the \ramindex is accessible starting from EL1, it appears
that it is gradually being elevated in the privilege level. For
instance, in Cortex-A77, EL3 (Trust-Zone security monitor level) is
required.

\section{\cachedump Overview}
\label{sec:overview}

\begin{figure}
  \centering
  \includegraphics[width=1\linewidth]{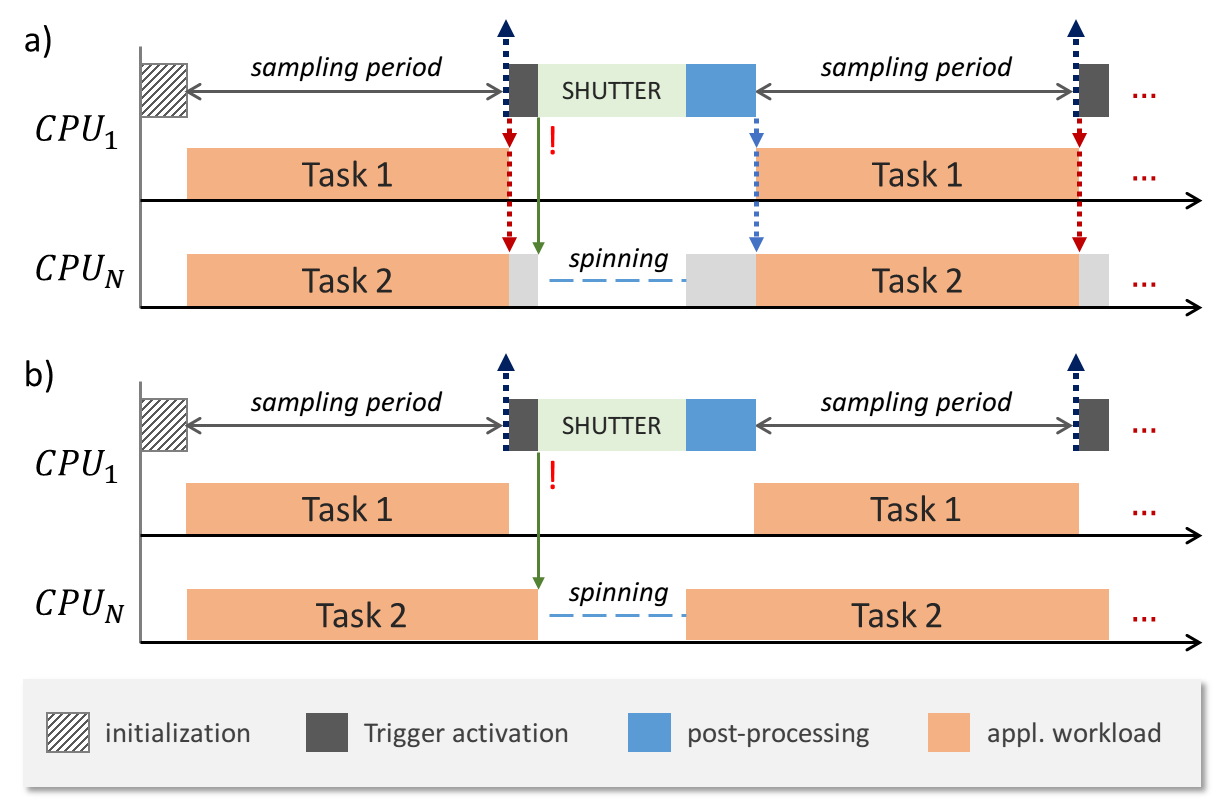}
  \caption{\trigger and \dumper modules operation over time in
    synchronous (a) and asynchronous (b) mode with periodic sampling.}
  \label{fig:timeline}
  \vspace{-0.5cm}
\end{figure}

In this section we discuss the general workflow of the proposed
\cachedump technique. We first provide a high-level description of the
different moving parts; we then describe the main challenges faced and
the possible usage scenarios for the proposed \cachedump.

\cachedump is structured in two modular components. The first module,
namely the \emph{\dumper} concerns the low-level logic that leverages
the \ramindex interface described in Section~\ref{sec:ramindex}. The
\dumper is responsible for initiating a snapshot of the content of the
target memory --- e.g. L1 data/tag, L2 data/tag, TLBs, etc. The
\dumper is implemented as an OS component and hence runs with
kernel-level privileges. It exposes two interfaces to the rest of the
system: (1) an interface to \rev{configure and initiate snapshot
  acquisition}; and (2) a data channel where the content of
\rev{acquired snapshots} is passed \rev{to user-space for final
  processing}.

The second module is the \emph{\trigger} that implements user-level
logic to commandeer a new snapshot to be performed by the \dumper. The
\trigger \rev{is} designed to support a number of event-based activation
strategies. For instance, to perform periodic sampling of the cache
content, the \trigger is activated via timer events.  Alternatively
the \trigger can be activated when the application under analysis
reaches a code or data breakpoint, invokes a system call, \rev{or
  delivers a specific POSIX signal to the \trigger process. Periodic
  activation, and event-driven activation initiated through signal
  delivery are currently implemented}.

\rev{
  \subsection{Flush vs. Transparent Mode}\label{sec:flush_vs_transparent_mode}
  \cachedump can operate in a number of modes to facilitate different
  types of cache analyses. A mode is selected by appropriately
  configuring the \dumper and \trigger modules. A more in-depth
  discussion is provided in Section~\ref{sec:impl}. Here, we
  provide a short overview of the most important modes. First, 
  \cachedump can operate in \emph{flush} or \emph{transparent} mode.
  When operating in flush mode, the
  acquisition of a snapshot is intentionally destructive to
  cache contents. In this mode, after acquiring a snapshot, the cache
  contents of the application(s) under analysis are flushed from the
  cache. Conversely, when operating in transparent mode, cache
  snapshotting is performed while minimizing the impact on the contents
  of the cache. We quantify the involuntary pollution overhead when
  operating in transparent mode in Section~\ref{sec:ev_overh}}.

\rev{
  \subsection{Synchronous vs. Asynchronous Mode}
  \label{sec:sync_vs_async_mode}
  \cachedump can also operate in \emph{synchronous} or
  \emph{asynchronous} mode. Synchronous mode is best suited to
  analyzing a specific subset of applications executing in parallel on
  multiple cores. In this mode, the \trigger spawns the applications
  under analysis and delivers POSIX signals to pause them once a new
  snapshot acquisition is initiated, and to resume them
  afterwards. Figure~\ref{fig:timeline}~(a) provides a timeline of events
  as they occur in the synchronous mode. When
  a new snapshot is to be acquired (dashed blue up-arrow in the
  figure), all the tasks under analysis are paused (dashed red
  down-arrows). Once the acquisition of the current snapshot is
  complete, all the observed applications are resumed (dashed blue
  down-arrows). This mode ensures that all the observed applications
  --- including the one executing on the same CPU as the \trigger ---
  are equally affected by the activation of the trigger. The extra
  complexity of pause/resume signals is unnecessary (1) when a single
  application is being observed, pinned to the same core as the
  \trigger; and (2) when one is not conducting an analysis on a
  specific set of applications, but, for instance, on the the
  \emph{background noise} of system services.  }

\rev{ To cover the latter two cases, \cachedump can operate in
  \emph{asynchronous} mode. In this case, there is no explicit
  pause/resume signal delivery to applications, as depicted in
  Figure~\ref{fig:timeline}~(b). Upon activation (dashed up-arrow),
  the \trigger preempts all the applications on the same core but does
  not explicitly pause applications on other cores. It then invokes
  the \dumper.}
  
\rev{Regardless of the mode, note that the \dumper temporarily spins
  all the cores (solid down-arrow) once invoked to perform the
  low-level interaction with \ramindex registers. This is necessary to
  ensure the correctness of the snapshot, as discussed in
  Section~\ref{sec:impl_dumper} and highlighted in
  Figure~\ref{fig:timeline}.}
  
  


\subsection{Key Challenges}
\label{sec:chall}

Three key challenges have been addressed in the proposed design of
\cachedump, which are hereby summarized. More details on how each
challenge was solved are provided in Section~\ref{sec:impl}.

\par{\bf Avoiding Pollution:} The first challenge we faced is quite
intuitive. Acquiring a snapshot of the cache involves the execution of
logic on the very same system we are trying to observe. Worse yet,
while the content of the cache is progressively read, one must use a
memory buffer to store the resulting data. But writes into the buffer
might trigger cache allocations and hence pollute the state of the
cache that is being sampled. Because the size of the used buffer needs
to be in the same order of magnitude as the size of the cache, this
issue can significantly impact the validity of the snapshot.

To solve this challenge, we statically reserve a portion of main
memory used by \cachedump to acquire a snapshot. The memory region
reserved for \cachedump is marked as non-cacheable. \rev{For
  \cachedump operating in flush mode, it is only necessary to reserve
  enough memory for a single snapshot. This space is reused for
  subsequent snapshots. Conversely, to operate in transparent mode,
  enough memory to store all the snapshots required for the current
  experiment needs to be reserved. We refer the reader to
  Section~\ref{sec:impl} for more details on the size of each
  snapshot.} \rev{With this setup, the \dumper minimizes snapshot
  pollution} by performing a loop of a few instructions that (1)
\rev{iterate} through all the cache ways and sets to be read; (2)
operate exclusively on CPU registers; and (3) only perform memory
stores toward the non-cacheable buffer.

\par{\bf Pausing Progress:} Capturing a snapshot can take a
non-negligible amount of time. While a snapshot capture is in
progress, it is important to ensure that the applications under
analysis do not progress in their execution. In other words, the
\dumper should be able to temporarily \emph{freeze} all the running
applications and resume their execution once the capture operation is
complete. Not doing so would result in snapshots that do not reflect a
real cache state. This is because the state of the cache would be
continuously changing while the capture is still in progress.

On a single-core implementation, it is enough to run the \dumper with
interrupts and preemption disabled to ensure that the application
under analysis does not continue to execute while a capture operation
is in progress. But this is not sufficient in a multi-core
implementation. To solve the problem, we first designate a master core
responsible for completing the capture operation. Next, we use
kernel-level inter-core locking primitives to temporarily stall all
the other cores. Once the snapshot has been acquired, all the other
cores are released and resume normal execution.

\par{\bf Inferring Content Ownership:} As recalled in
Section III, shared caches --- like the L2 targeted
in our implementation --- are generally PIPT. As such, when a snapshot
is captured, we obtain a list of physical address tags. A first
important step consists in reconstructing the full physical address
given the obtained tag bits and the cache index bits used to retrieve
each tag. The end goal of our analysis, however, is to attribute
captured cache lines to running applications or OS-level
components. This step is strictly dependent in the strategy used by
the OS/platform under analysis to map applications' virtual addresses
to physical memory. We distinguish three cases.

The first and simplest case corresponds to (RT)OS's operating on small
micro-controller that do not have support for virtual memory,
i.e. where no MMU is present. These systems usually feature a Memory
Protection Unit (MPU) that allows defining permission regions for
ranges of physical addresses. Both applications and OS components are
then directly compiled against physical memory addresses. In this
case, ownership of cache blocks can be inferred by simply comparing
the obtained physical addresses w.r.t. the global system memory map.

The second case corresponds to systems where, albeit an MMU exists, it
is configured to perform a flat linear mapping between virtual
addresses and physical addresses. In this case there exists a
(potentially null) constant offset between virtual addresses and
corresponding physical addresses. 

The third scenario corresponds to OS's that use demand paging. In this
case, there is no fixed mapping between virtual pages assigned to
applications and physical memory. In this case, contiguous pages in
virtual memory are arbitrarily mapped to physical memory, following
the OS's internal memory allocation scheme. With demand paging,
applications are initially given a virtual addressing space. Only when
the application ``touches'' a virtual page, a new physical page is
allocated from a pool of free pages. In \cachedump we consider this
case because it represents the most general and challenging scenario.

\subsection{Usage Scenarios}\label{sec:usage_scenarios}
We envision that, in addition to the use cases directly explored in
this paper, \cachedump and variations of our technique can be employed
in a number of use cases, including but not limited to the
following: (1) To study the heat-map of cached content over time and
understand the evolution of the active memory working set of user
applications. (2) To study the cache footprint of specific functions
within an application, a library or at the level of system calls. (3)
In a single-core setup with multiple co-running applications, to study
how scheduling decisions and preemptions affect the composition of
cache content over time. (4) In a multi-core setup with multiple tasks
running in parallel to study the contention over cache resources. (5)
To debug and validate set-based (e.g. coloring) and way-based cache
partitioning techniques. (6) To validate hypotheses on the expected
behavior of a multi-level cache hierarchy, e.g., in terms of
inclusiveness, replacement policy, employed coherency protocol,
prefetching strategies. (7) To asses the vulnerability of a machine to
cache-based side channel attacks arising from speculative execution
and memory traffic.

\subsection{Comparison to Other Approaches}\label{sec:comparison_hw}
\cachedump offers a novel method to study caches, where traditionally
hardware debuggers and simulation models have been used.
System designers have traditionally resorted to hardware debuggers to inspect the
contents of cache memories, using them as a proxy to study correct
system behavior and explain applications' performance. The main advantage
of using an external hardware debugger to inspect the state of caches is that the
impact of the debugger on the cache itself can be kept to a minimum.
But making sense out of a cache snapshot requires access to OS-level
data structures such as page tables and VMA layouts, to name a few.
Debuggers that provide some of the cache analysis features provided by \cachedump rely on
high-bandwidth trace ports --- as opposed to traditional JTAG ports ---
often unavailable in production systems.
The Lauterbach PowerTrace II and the ARM DS-5 with the DSTREAM
adapter are examples of solutions that can provide snapshots of cache contents.
Their price tag exceeds USD 6,000 
\footnote{See \url{http://www.wg.com.pl/pliki/cennik/2017\%20Prices\%20DS-5.pdf}}.
While in principle an inexpensive JTAG debugger could be used to halt
the CPUs, interact with the \ramindex interface, perform
physical$\rightarrow$virtual translation and application layout
resolution, no such implementation exists to the best of our
knowledge. 


In contrast, \cachedump runs entirely in software, imposes minimal
system overhead, does not requires the existence of a debug port nor extra hardware,
and can run on most machines with support for
the \ramindex interface, while managing to provide much of the same
information as hardware debuggers with minimal effort. \cachedump's
most obvious shortcoming is some inevitable overhead, in terms of cache pollution,
compared to hardware debuggers\footnote{We chose to implement \cachedump as a Linux
kernel module for flexibility. It is theoretically possible, however,
to rewrite the module to run at a lower level
(i.e. at the hypervisor level).}.

\par Another method often used to perform cache profiling is
simulation.  Unfortunately, simulation models are often too generic to
capture implementation-specific design choices.
Gem5~\cite{gem5_2011}, for example, only simulates a generic cache
model, which may not be in match with the behavior of the actual
hardware. It is also challenging to simulate entire systems with
production-like setups in terms of active system services, active I/O
devices and concurrent applications. Conversely, \cachedump can be
used to observe the behavior of a system in the field, and/or to
validate and refine platform-specific cache simulation models.

\par Yet another class of cache analysis approaches are based on
performance-counter sampling. These only provide quantitative
information on system-wide metrics that are best interpreted with a
good understanding of the micro-architecture at hand.  In comparison,
\cachedump, provides behavioral information about the cache that is
akin to what a hardware debugger could provide.

An additional benefit \cachedump offers compared to the aforementioned
approaches is its relative versatility apropos deployability. Since it
relies exclusively on \ramindex and Linux's scaffolding for building
and loading kernel modules, \cachedump serves as an excellent
candidate for remote deployment. On virtual private servers (VPS), for
instance, \cachedump can provide information that developers would
traditionally rely on debuggers for, without necessitating physical
access to the system. As such, \cachedump's value lies primarily in
its simplicity and its reliance on ubiquitous support structures, both
of which then engender an acceptable compromise between the effort
needed to setup a hardware debugger and the loss of granularity
incurred when using simulators.

\section{Implementation}
\label{sec:impl}

\begin{figure}
  \centering \includegraphics[width=\linewidth]{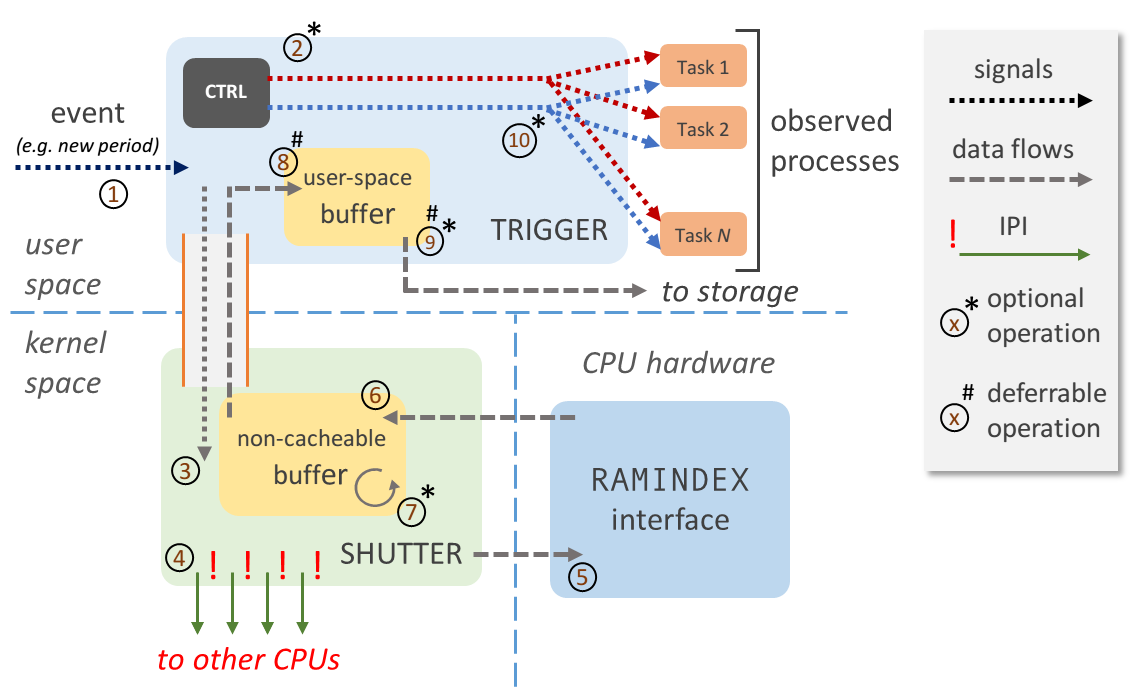}
  \caption{Logical interplay between the modules of \cachedump and
    sequence of operations performed to capture a cache snapshot.}
  \label{fig:steps}
  \vspace{-0.5cm}
\end{figure}

\newcommand{\step}[1]{Figure~\ref{fig:steps}\circled{#1}\xspace}

Additional details about our \cachedump implementation follow below.
We begin by describing the relevant features of our target SoC and
then illustrate the workings of the \trigger and \dumper modules.
An open-source version of \cachedump is available at: \textit{
URL omitted for blind submission.}

\subsection{Target Platform}
\par We conducted our experiments on an NVIDIA Tegra X1
SoC~\cite{tx1-trm}. The TX1 chip features a cluster of four ARM Cortex
A57~\cite{arm-cortex-a57} CPUs operating at a frequency of 1.9~GHz,
along with four unused ARM Cortex-A53 cores. Each CPU contains a
private 48~KB L1 instruction cache and a 32~KB L1 data cache. The
L2\textemdash also the LLC\textemdash is unified and shared among all
the cores.  It is implemented as a PIPT cache and employs a random
replacement policy.

In terms of geometry, the L2 cache is $C_S = $ 2~MB in size. The line
size is $L_S = $ 64~bytes and the associativity is $W = $ 16. It
follows, then, that the cache is divided into 2048 cache sets\textemdash each
containing 16 cache lines, which in turn contain 64~bytes of data each. 
Bits [0, 5] of a physical address mark the offset bits;
bits [6, 16] are the index bits; bits [17, 43]\footnote{The platform
  supports a 44-bit physical address space.} correspond to the tag
bits.

\rev{The information acquired for each cache line in our
  implementation is limited to 16~bytes. Of these, 8~bytes are for the
  PID of the process that \emph{owns} the line, and the remaining
  8~bytes encode an address field. If address resolution is turned on,
  the field holds the resolved virtual address. If address resolution
  is turned off, the field is used to store the raw physical address
  instead. For a 16-way set-associative cache with a line size of
  64~bytes and total size of 2~MB, like the one considered for our
  evaluation, a single snapshot is 512~KB in size. In our setup, we
  have dedicated 1~GB of memory for \cachedump, meaning that for any
  given experiment, up to 2048 snapshots can be collected. With a
  typical snapshot period of 5-10ms, this allows studying the behavior
  of applications with a runtime of 10-20 seconds.}

\subsection{\trigger Implementation}
Starting from the top-level module of \cachedump, i.e. the \trigger,
we hereby review the proposed implementation following the logic flow
of operations provided in Figure~\ref{fig:steps}.

The \trigger module is always executed with the highest real-time
priority. The module is designed for event-based activation and hence
a new snapshot is initiated when a new event activates the \trigger,
as shown in \step{1}. In our current prototype, \rev{we implemented
  two activation modes: (1) periodic activation with a configurable
  inter-activation time; and (2) event-based. In periodic mode,}
events are generated using a real-time timer set to deliver a
\texttt{SIGRTMAX} signal to the \trigger. The tasks under analysis are
launched directly by the \trigger to ensure in-phase snapshotting, and
to allow the \trigger to set a specific scheduler and priority on the
spawned children processes.

\rev{The \trigger implements event-based activation by initiating a
  snapshot upon receipt of a \texttt{SIGRTMAX-1} signal. In
  synchronous mode, a task under analysis spawned by the \trigger can
  initiate a snapshot with a combination of \texttt{getppid} and
  \texttt{kill} system calls. This mode was used in
  Section~\ref{sec:ev_repl} to study the properties of the cache
  replacement policy in the target platform. The limitation of this
  approach is that some instrumentation in the applications' code is
  required.} Future extensions will leverage the \texttt{ptrace}
family of operations to allow attaching to unmodified applications,
and to trigger a new snapshot upon reaching an instruction or data
breakpoint.

The next step depends on the mode (synchronous vs. asynchronous) in
which the \trigger is configured to run. In the \emph{synchronous}
mode, the trigger immediately stops all the observed tasks with a
\texttt{SIGSTOP} --- see \step{2}. This is particularly useful if
multiple cores are active, but introduces unnecessary additional
overhead when performing single-core analysis. Hence, this is an
optional step and skipped when the \trigger operates in
\emph{asynchronous} mode.

Next, the \trigger commandeers the acquisition of a new snapshot to
the \dumper module via a \texttt{proc} filesystem interface, as
depicted in \step{3}. \rev{If the trigger is operating in flush mode,}
the binary content of the snapshot is immediately copied to a
user-space buffer via the same interface, as shown in \step{8}. At
this point, the trigger might collect in-line statistics on the
content of the snapshot, or (optionally) (see \step{9}) render the
snapshot in human-readable format and store it in persistent memory
for later analysis. \rev{This step is deferred to the end of the
  experiment for all the collected snapshots if \cachedump operates in
  transparent mode.}

\rev{Typical embedded applications make limited use of dynamic memory
  and dynamically linked libraries, which are instead commonly used
  features in general-purpose applications. It follows that the
  virtual memory layout --- i.e. the list of VMAs --- of embedded
  applications is generally static. It is easy to infer to which VMA a
  given virtual address belongs to in applications with static memory
  layout. Conversely, dynamically linking libraries and
  allocating/freeing memory at runtime can substantially change the
  VMA layout of an application. For this reason, the \trigger
  optionally allows recording the current memory layout of an
  application at the time a cache snapshot is acquired. This is done
  by reading the \texttt{/proc/PID/maps} interface, where \texttt{PID}
  is the process id of the application under analysis.}

Finally, if the \trigger is operating in synchronous mode, a
\texttt{SIGCONT} is sent to all the tasks under analysis as shown in
\step{10}.



\subsection{\dumper Implementation}
\label{sec:impl_dumper}

The \dumper is implemented as a Linux kernel module. At startup, it
establishes a communication channel with user-space by creating a new
entry in the \texttt{proc} pseudo-filesystem. \rev{Configuration
  parameters and snapshot acquisition commands are sent to the \dumper
  via \texttt{ioctl} calls. If \cachedump is operating in transparent
  mode, this interface is also used to specify which snapshot to
  retrieve at the conclusion of the experiment. A \texttt{read} system
  call (\step{3}) can be used to retrieve the selected snapshot in
  user-space.}

\par{\bf Inter-CPU Synchronization:} As discussed in
Section~\ref{sec:overview}, being able to perform a full capture of
L2's content while minimizing pollution is of the utmost
importance. For this reason, the \dumper delivers an Inter-Processor
Interrupt (IPI) to all the other CPUs, forcing them to enter a
busy-waiting loop --- see \step{4}. Specifically, right before
broadcasting the IPI, the \dumper acquires a spinlock. Next, the
payload of the delivered IPI makes all the other CPUs wait on the
spinlock.

\par{\bf Pollution-free Retrieval:} While holding the spinlock, the
\dumper proceeds to use the aforementioned \ramindex (see
Section IV) interface to access the contents of the L2
tag memory, as shown in \step{5}. Entry by entry, a kernel-side buffer
is filled with the result of the \ramindex operations --- as per
\step{6}. To avoid polluting the L2 at this step, the buffer is
allocated as a non-cacheable memory region. To do so, we use the
boot-time parameter \texttt{mem} to restrict the amount of main memory
seen (and used) by Linux to carve out a large-enough
physically-contiguous memory buffer. This area is then mapped by the
\dumper using the \texttt{ioremap\_nocache} kernel API and used to
receive the content of the L2 tag entries.

\par{\bf From Physical to Virtual:} After all the L2 tag entries have
been retrieved, the buffer contains a collection of physical
addresses, one per each cache block that was marked as \emph{valid} in
L2. Recall from Section~\ref{sec:chall} that operating systems that
have full support for MMUs \rev{define} a non-linear mapping between
virtual pages and physical memory such that a set of contiguous
virtual pages is comprised of arbitrarily scattered physical
pages. 
Therefore, while the conversion virtual$\rightarrow$physical can be
easily performed using page-table walks, the reverse translation is
non-trivial. The \rev{physical address resolution} step depicted in
\step{7} refers to such a reverse translation performed on each of the
retrieved L2 entries.

To perform this \rev{resolution}, we leverage Linux's specific
representation of memory pages. Linux defines a descriptor of type
\texttt{struct page} for each of the physical memory pages available
in the system. \rev{The conversion fom physical address to page
  descriptor is possible through the \texttt{phys\_to\_page} kernel
  macro}. We first \rev{derive the page descriptor of the physical
  address to be resolved.} Next, we effectively re-purpose the
reverse-map interface\footnote{See
  \url{https://lwn.net/Articles/75198/} for more details.}
(\texttt{rmap}) used by Linux to efficiently free physical memory when
swapping is initiated. The entry point of the interface is the
\texttt{rmap\_walk} kernel API. Given a target \texttt{struct page}
descriptor, the procedure allows one to specify a callback function to
be invoked when a possible candidate for the reverse translation is
found\footnote{Because multiple processes might be mapping the same
  physical memory, the reverse translation is not always unique.}. A
successful \texttt{rmap\_walk} operation returns (i) a reference to
the VMA that maps the page; and (ii) the virtual address of the page
inside the VMA. Importantly, the reference to the VMA allows one to
derive the original memory space (\texttt{struct mm\_struct}); and
from there, the descriptor of the process (\texttt{struct
  task\_struct}) associated to the memory space and its PID. After the
translation step, the entries in the buffer are converted to contain
two pieces of information: (i) the PID of the process to which the
cache block belongs, and (ii) the virtual address of the block within
the process' virtual memory space.

\rev{The virtual$\rightarrow$physical translation can be optionally
  disabled. This is useful, for instance, when profiling the cache
  behavior of an application pinned to a specific subset of physical
  pages, of a different virtual machine, or of the kernel itself.}

\par{\bf From Kernel to User:} Lastly, the contents of the kernel-side
buffer are copied into a user-space buffer defined in the
\trigger. \rev{On this very last step, a distinction needs to be made
  because the behavior of \cachedump significantly differs when it
  operates in flush mode, compared to transparent mode operation.}

\rev{In flush mode, the goal is to analyze what cache blocks are
  actively loaded by an application in between snapshots. For this
  reason, every snapshot acquisition is immediately followed by a copy
  of the snapshot to the \trigger in user-space\footnote{Note that it
    is still important to prevent cache pollution \emph{while} the
    current snapshot is being acquired. Thus, pollution-free retrieval
    is still crucial.}. The \trigger also converts the binary format
  of the snapshot to human-readable format and writes it to disk.
  This step corresponds to \step{8}. The copy to user-space, as well
  as any post-processing performed by the \trigger, is conducted in
  cacheable memory. Because the amount of data moved after each
  snapshot is comparable in size to the L2, the post-processing acts
  as a tacit \emph{flush} operation. However, to cope with random
  cache replacement policies, the \trigger performs additional cache
  trashing before resuming the applications to ensure that the content
  of the cache is indeed flushed.} The presence of this flush
operation is vital to the correct interpretability of the results. By
doing so, each snapshot contains only cache blocks allocated during
the last sampling period. Therefore, the extracted content is
representative of the recent activity of the applications and enables
active cache working-set analysis.

\rev{In transparent mode, the goal is to analyze the evolution of
  cache content over time while minimizing the impact of \cachedump on
  the cache state. In this mode, no post-snapshot flush is
  performed. Thus, subsequent snapshots are accumulated in
  non-cacheable memory. They are then moved to user-space and
  post-processed only at the very end of the experiment. We evaluate
  in Section~\ref{sec:ev_overh} how much pollution is introduced
  in the various modes of operation.}

\rev{Because the size of a snapshot to be transferred to user-space is
  on the order of hundreds of pages, we use} the sequential file
(\texttt{seq\_file}) kernel interface\footnote{See
  \url{https://lwn.net/Articles/22355/} for more details.}. This
interface safely handles \texttt{proc} filesystem outputs spanning
multiple memory pages.

\section{Evaluation}
\label{sec:eval}

\newcommand{\bmseq}{\textsc{Synth}\xspace}
\newcommand{\bmstep}{\textsc{SynthStep}\xspace}
\newcommand{\disp}{\textsc{Disparity}\xspace}
\newcommand{\mser}{\textsc{Mser}\xspace}
\newcommand{\sift}{\textsc{Sift}\xspace}
\newcommand{\track}{\textsc{Track}\xspace}
\newcommand{\bomb}{\textsc{Bomb}\xspace}
\newcommand{\repl}{\textsc{Repl}\xspace}

This section aims to demonstrate the capabilities of the proposed and
implemented \cachedump technique. This is not meant to be an
exhaustive evaluation of all the scenarios in which \cachedump might
be employed, but rather to demonstrate that \cachedump is capable of
producing useful insights on the cache usage of real applications in a
real system.

\subsection{Setup and Goals}
All the experiments described in this section have been carried out on
an NVIDIA Jetson TX1 development system running Linux v4.14. The
Jetson TX1 features an NVIDIA Tegra X1 SoC, in line with what is
described in Section~\ref{sec:impl}.

For our workload, we use a combination of synthetic and real
benchmarks. \rev{Additional details about the synthetic benchmarks we
  designed are provided contextually to the experiment in which they
  are employed. For our real benchmarks, we considered} applications
from the the San Diego Vision Benchmarks (SD-VBS)~\cite{sd-vbs}, which
come with multiple input sizes. Our goal is to demonstrate the
usefulness of \cachedump in analyzing an application's cache
behavior. As such, we include only a selection of the obtained results
covering the most interesting cases. 
We selected the \disp, \mser, \sift, and \track
benchmarks with intermediate input sizes, namely VGA (640x480) and CIF
(352x288) images.


The remainder of this section is organized to address the following
questions: \begin{enumerate}
\item Is \cachedump able to provide an output that is representative
  of the actual cache behavior of an application? This is covered in
  Section~\ref{sec:ev_synth}.
\item \rev{What is the overhead in terms of cache content pollution
  and time?} We discuss this aspect in Section~\ref{sec:ev_overh}.
\item Is it possible to track the cache behavior of real applications
  in term of WSS and frequently accessed memory locations using
  \cachedump? We tackle this question in Section~\ref{sec:ev_wss}
\item Can \cachedump reveal system-level properties, such as (1) how
  the cache is being shared by concurrent applications; (2) how
  scheduling decisions impact cache usage? Section~\ref{sec:ev_sys}
  approaches this question.
\item \rev{Is it possible to use \cachedump to predict whether
  an application will suffer measurable cache interference from a
  co-running application? This is explored in
  Section~\ref{sec:ev_compat}.}
\item \rev{Can we study the replacement policy and implemented by the
  target platform and its statistical characteristics? This analysis
  is conducted in Section~\ref{sec:ev_repl}.}
\end{enumerate}


\subsection{Is \cachedump's Output Meaningful?}
\label{sec:ev_synth}

The first aspect to validate is whether or not \cachedump is able to
provide an output that can be \emph{trusted}, in the sense that it is
meaningfully related to the behavior of the application under
analysis. We \rev{study the output produced by \cachedump on a
  synthetic benchmark. The benchmark allocates two buffers of 512~KB
  each. It then performs a full write followed by a full read on the
  first buffer. Next, it performs a full write followed by a full read
  on the second buffer.}

\begin{figure}
  \centering
  \includegraphics[trim={0 0.8cm 0 0.2cm}, clip, width=\linewidth]{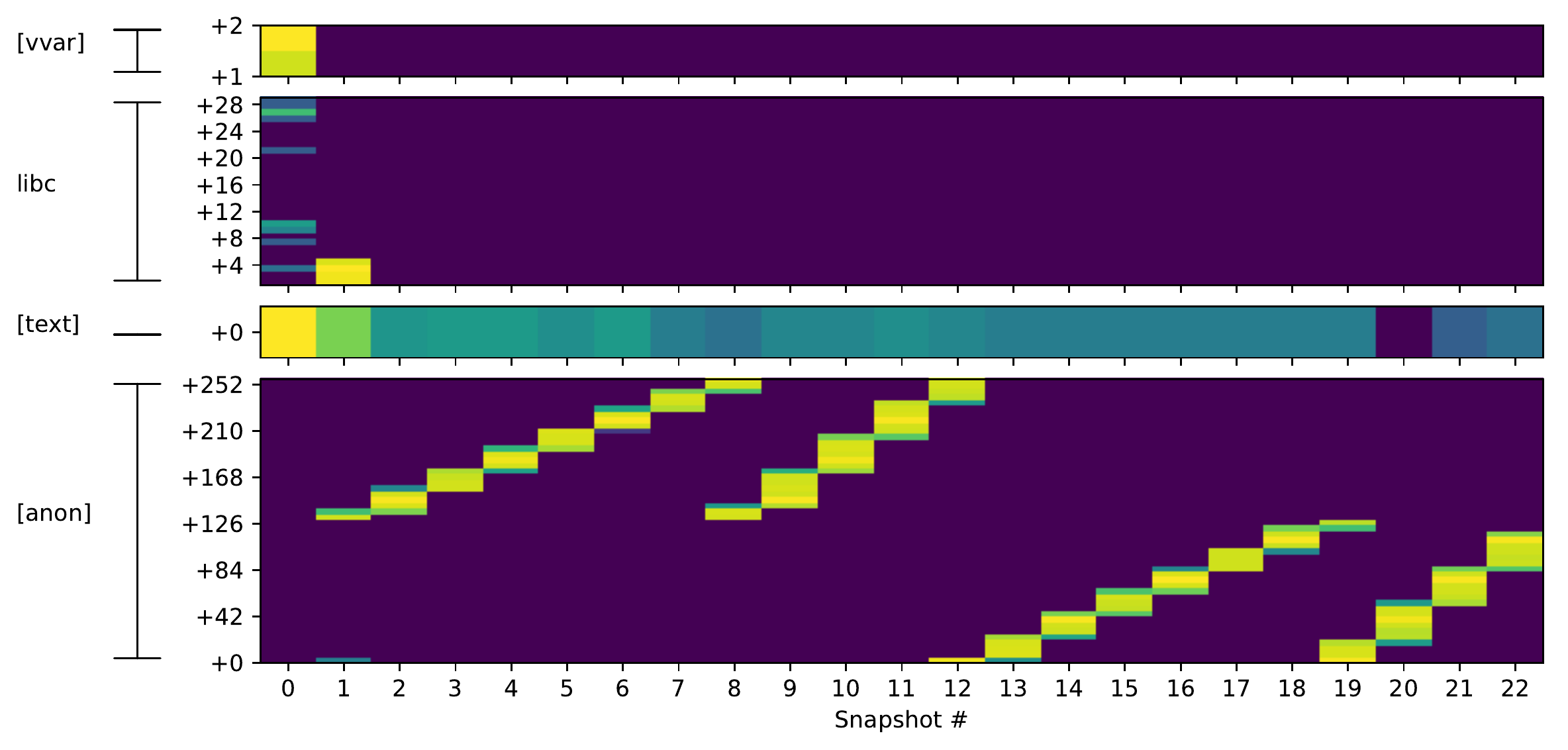}
  \vspace{-0.7cm}
  \caption{\bmseq memory heat-map analysis. Flush mode.}
  \label{fig:synth_fl}
  \vspace{-0.4cm}
\end{figure}

\begin{figure}
  \centering
  \includegraphics[trim={0 0.8cm 0 5cm}, clip, width=\linewidth]{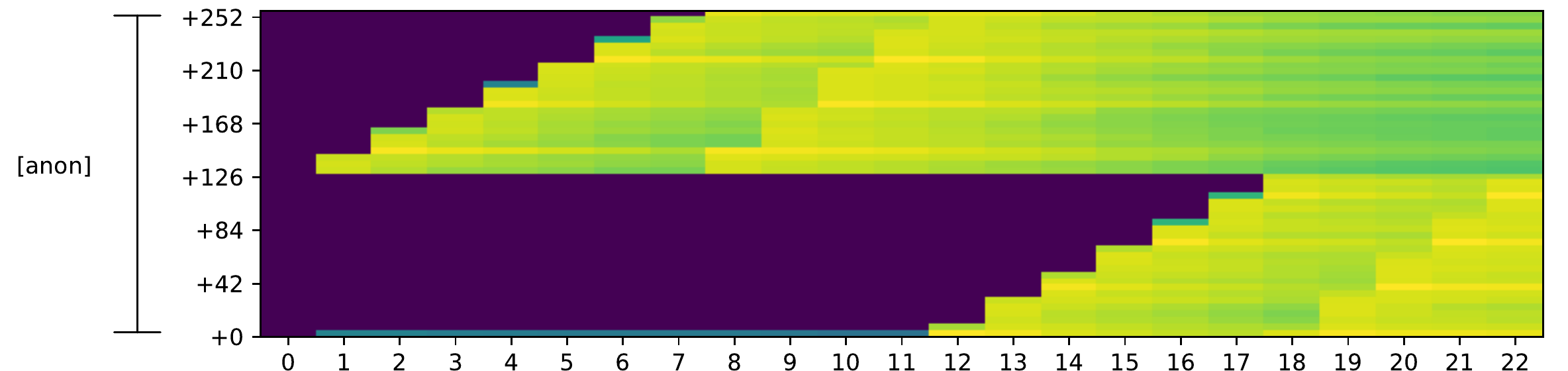}
  \vspace{-0.7cm}
  \caption{\bmseq memory heat-map analysis. Transparent mode.}
  \label{fig:synth_tr}
  \vspace{-0.4cm}
\end{figure}

We set our trigger to operate in periodic, synchronous mode, with an
interval of \rev{2}~milliseconds between snapshots. We then plot a
heat-map of the number of cache lines found in the snapshot for each
of the pages that belong to the benchmark under analysis. \rev{We
  perform the experiment twice, once in flush mode and then in
  transparent mode. The results of the two experiments are depicted in
  Figure~\ref{fig:synth_fl} and~\ref{fig:synth_tr}.}

In Figure~\ref{fig:synth_fl} we depict the heat-map for the 4 most
used memory regions --- as determined by reading the content of the
snapshots. We only depict the most used \rev{(and interesting)} region
in Figure~\ref{fig:synth_tr}. The color scale we used indicates the
number of lines present per 4~KB page in each snapshot, with darker
blue tones indicating fewer lines and tones shifting towards yellow
signifying more lines. The progression of the snapshots is plotted on
the $x$ axis. Page addresses are then plotted on the $y$ axis in terms
of relative offset (in pages) from the beginning of the region. We
annotate the name of the region on the left-hand side of the plot.

A number of important observations can be made. We summarize the most
interesting ones. First, we notice that at the beginning of the
application, the pages of the stack, the text and \texttt{glibc}
library are mostly present in cache. This is in line with the
initialization of the application which uses \texttt{malloc} to
allocate its buffer. For small allocations, \texttt{malloc} expands
the heap of the calling process. But if the allocation of larger
buffers is requested, the \texttt{glibc} resorts to performing a
\texttt{mmap} call instead to create a new memory region not backed by
a file (anonymous memory). It is in this anonymous region, marked as
``[anon]'' in the plot, that the bulk of memory accesses will be
performed by our benchmark. We can also notice that the region
comprises \rev{256 pages, i.e. 1~MB}.

Let us now focus on this region. In Figure~\ref{fig:synth_fl}, one can
easily distinguish the \rev{scan performed by the benchmark over the
  buffers}. \rev{Each scan} shows up as bands of yellow moving
left-to-right. Because a full pass of stores followed by a full pass
of loads is performed in each \rev{scan}, two \emph{yellow bands} are
visible in each \rev{scan}. By looking at the slope of these bands, it
is possible to understand the rate of progress of the benchmark moving
through the buffer. It can be noted that stores are slower than loads
--- it takes around 8 snapshots to complete the first store
sub-iteration, and only 5 for the loads. This might sound
counter-intuitive but indeed makes sense because (1) in write-back
caches a store might result in a write-back of a dirty line followed
by a load from main memory; and (2) a cache with many outstanding
store transactions will stall when its internal write-buffer is full.

\rev{Because Figure~\ref{fig:synth_fl} was produced in flush mode, the
  only cache lines highlighted in the heat-map are those that were
  accessed by the application in-between snapshots. It goes to
  demonstrate that flush mode is particularly well suited to study the
  active cache set of applications. Conversely, operating in
  transparent mode allows us to understand how/if cache lines allocated
  at some point during execution persist in
  cache. Figure~\ref{fig:synth_tr} depicts one such case. After
  snapshot 12, the application will not access the top portion of its
  addressing space. Nonetheless, since the overall buffer touched by
  the application is smaller than the cache size, the unused lines
  remain in cache. These lines slightly fade in
  color because some eviction still occurs due to the random
  replacement policy of this cache.}

\subsection{\rev{What is the Overhead of Snaphotting?}}
\label{sec:ev_overh}

\rev{We evaluate the overhead introduced by \cachedump along
  two dimensions: cache pollution and timing.
  We measure cache pollution as the
  change in cache content that is introduced solely because of
  \cachedump's activity. To evaluate cache pollution, 
  we first execute a cache-intensive application
  for a given \emph{lead} time set to 100~ms. The lead time allows the
  application to populate the cache.  Then, we set the \trigger to activate
  at the end of the lead time and to acquire two consecutive snapshots
  back-to-back. The first snapshot captures the content of the cache
  status as populated by the application under analysis. The second
  snapshot is used to understand the change in cache status introduced
  by kernel-to-user copy of the first snapshot and its
  post-processing. The overhead then is evaluated in terms of
  percentage of cache lines that have changed over the total number of
  cache lines. The overhead in time is evaluated by measuring the
  end-to-end time required to acquire a single snapshot.}

\begin{table}[]
  \caption{\rev{Pollution and Time overhead of \cachedump.}}
  \resizebox{\columnwidth}{!}{%
  \begin{tabular}{l|cccccc}
    & \multicolumn{3}{c|}{\textbf{Space Overhead (\%)}}                                                             & \multicolumn{3}{c}{\textbf{Time Overhead ($10^6$ Cycles)}}   \\
    \textbf{Mode}             & \multicolumn{1}{l}{\textbf{Avg.}} & \multicolumn{1}{l}{\textbf{Std. Dev}} & \multicolumn{1}{l|}{\textbf{Max}} & \textbf{Avg.} & \textbf{Std. Dev.} & \textbf{Max} \\ \hline \hline 
    \textit{Full Flush}       & 95.23                             & 3.21                                  & \multicolumn{1}{r|}{99.10}        & 10.13 & 0.78 & 12.88 \\ \hline
    \textit{Resolve+Layout}   & 13.40                             & 0.43                                  & \multicolumn{1}{r|}{14.70}        & 0.34 & 0.00  & 0.35  \\ \hline
    \textit{Resolve}          & 6.78                              & 0.45                                  & \multicolumn{1}{r|}{8.23}         & 0.34 & 0.00  & 0.34  \\ \hline
    \textit{Layout}           & 6.55                              & 0.50                                  & \multicolumn{1}{r|}{8.71}         & 0.20 & 0.00  & 0.20  \\ \hline
    \textit{Full Transparent} & 1.06                              & 0.19                                  & \multicolumn{1}{r|}{1.38}         & 0.19 & 0.00  & 0.19
  \end{tabular}
  }
  \label{tab:overhead}
  \vspace{-0.5cm}
\end{table}

\rev{The results of the overhead measurements are reported in
  Table~\ref{tab:overhead} and are the aggregation of 100 experiments
  for each setup. We measure pollution and time overhead in
  synchronous mode because it is the most suitable for general
  analysis scenarios. We then consider a host of different
  sub-modes. In ``\emph{Full Flush}'', \cachedump operates in flush
  mode with physical$\rightarrow$virtual translation and VMA layout
  acquisition active. As expected, around 95\% of the cache content is
  modified in this mode after a snapshot is completed. The next four
  cases correspond to the overhead of \cachedump operating in
  transparent mode. In the ``\emph{Resolve+Layout}'' mode, both
  address resolution and VMA layout acquisition are turned on. In the
  ``\emph{Resolve}'' (resp., ``\emph{Layout}'') case, only address
  resolution (resp., VMA layout acquisition) are enabled. Finally in
  the ``Full Transparent'' case, both address resolution and VMA
  layout acquisition are skipped. Notably, cache pollution in full
  transparent mode does not exceed 1.4\%, which is acceptable for a
  software-only solution. Times are reported in millions of CPU cycles
  to better generalize to other platforms. The CPUs operate at 1.9~GHz
  on the target platform.}

\subsection{Can \cachedump Analyze Real Applications?}
\label{sec:ev_wss}
\begin{figure}
  \centering
  \includegraphics[width=\linewidth]{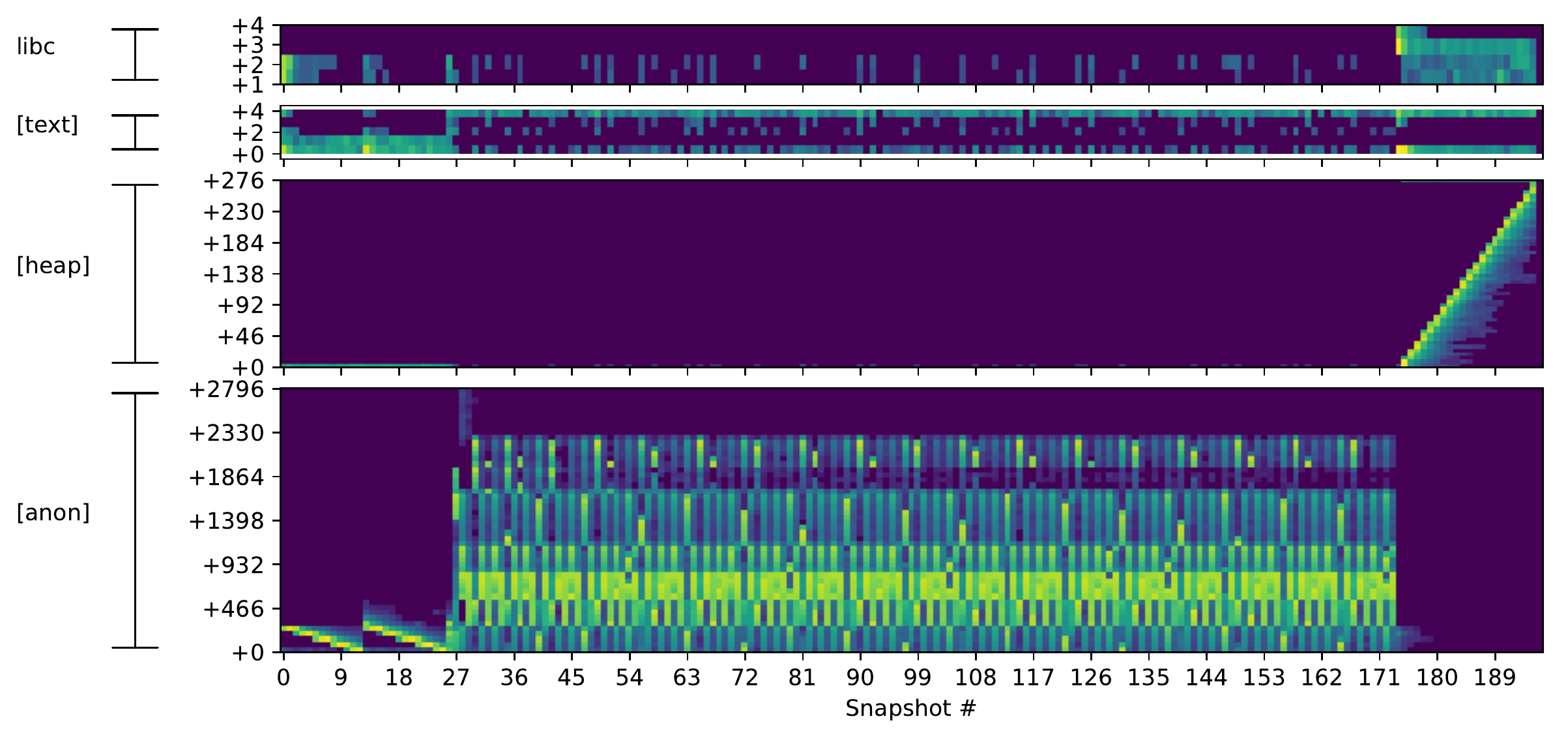}
  \vspace{-0.7cm}
  \caption{\disp memory heat-map when executed on VGA input.}
  \label{fig:disp_vga}
  \vspace{-0.4cm}
\end{figure}

\begin{figure}
  \centering
  \includegraphics[width=\linewidth]{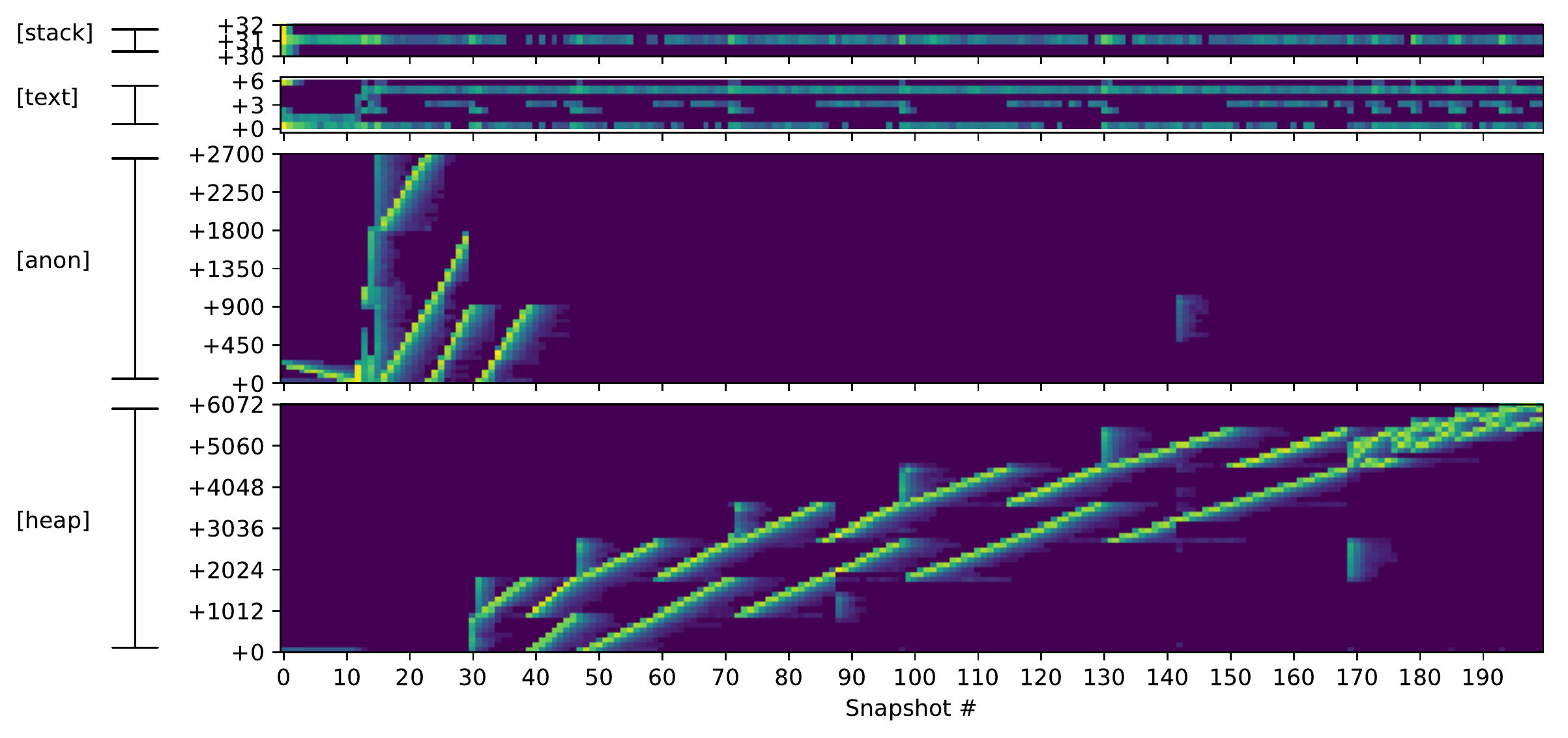}
  \vspace{-0.7cm}
  \caption{\sift memory heat-map when executed on VGA input.}
  \label{fig:sift_vga}
  \vspace{-0.4cm}  
\end{figure}

Having assessed that \cachedump can indeed provide accurate insights
into an application's cache utilization, we analyze two real
applications. Figure~\ref{fig:disp_vga} and Figure~\ref{fig:sift_vga}
report the heat-map analysis of SD-VBS benchmarks \disp and \sift,
respectively, using VGA resolution images in input.

It can be observed that \disp is characterized by quite
distinguishable \emph{intro} (snap.~0-25) and \emph{outro} phases
(snap.~172-198). In the intro, the input buffer is first pre-processed
in the ``anon'' region. It follows an intermediate processing phase
that actively uses around 83\% of the region's memory with a recurring
pattern and pages with offset 466-930 being the most frequently
accessed. During the outro phase the final output is produced
sequentially on the heap. A quite different picture is painted by the
\sift application. In this case, there exists an initial phase
(snap.~0-40) where pre-processing is performed on an anonymous region;
then the bulk of processing is carried out on the heap. From around
snapshot 30 until 170, two non-contiguous sets of memory pages are in
use, that gradually span through 82\% of the region. The final
computation step is performed on the top 75\% of the heap.

\subsection{Can \cachedump Discover System Properties?}
\label{sec:ev_sys}

\cachedump can also provide valuable insights on the cache-related
behavior of the whole system when multiple applications are
executed. \rev{To evaluate this,} we concurrently run four SD-VBS
applications, namely \track, \mser, \sift, \disp, all with VGA
resolution images in input.

\begin{figure}
  \centering
  \includegraphics[width=\linewidth]{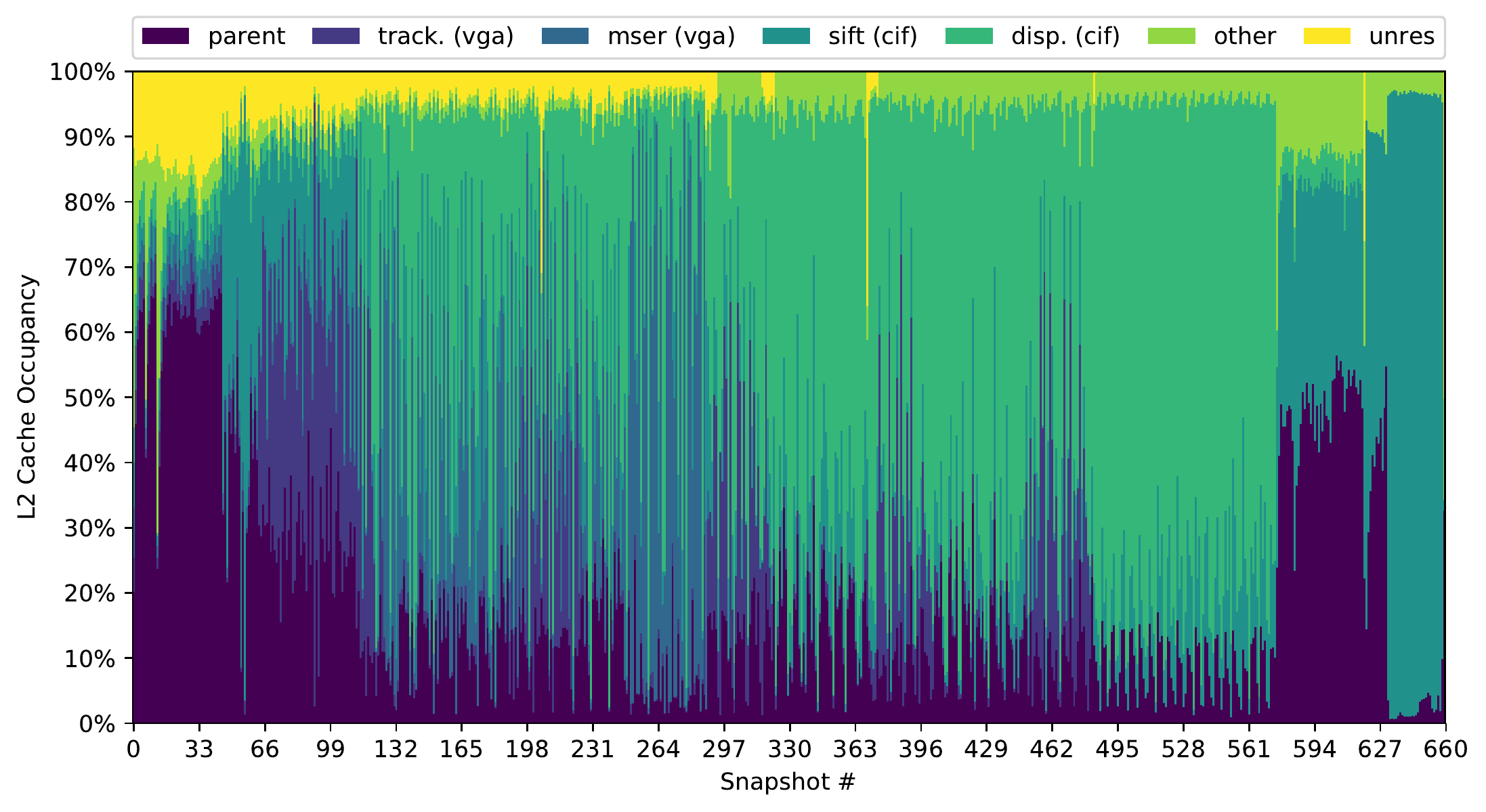}
  \vspace{-0.7cm}
  \caption{Single-core execution of 4 SD-VBS benchmarks with default
    completely fair scheduler (Linux's default).}
  \label{fig:4sdvbs_nrt}
  \vspace{-0.4cm}
\end{figure}

In the first experiment, the benchmarks are executed on a single core
--- the other three cores are turned off. Moreover, the default
Linux's scheduling policy (\texttt{SCHED\_OTHER}) is used. We then set
the Trigger to acquire periodic snapshots every 10~ms and study the
per-process L2 occupancy over time. The results are shown in
Figure~\ref{fig:4sdvbs_nrt}. The \texttt{SCHED\_OTHER} scheduler
implements a Completely Fair Scheduler (CFS), which tries to ensure an
even progress of co-running workload. This results in frequent
context-switches that result in quick changes in the composition of
the L2 cache content visible in the figure. The figure also shows how
more cache-intensive benchmarks such as \disp can dominate other
workload in terms of utilization of cache resources.

\begin{figure}
  \centering
  \includegraphics[width=\linewidth]{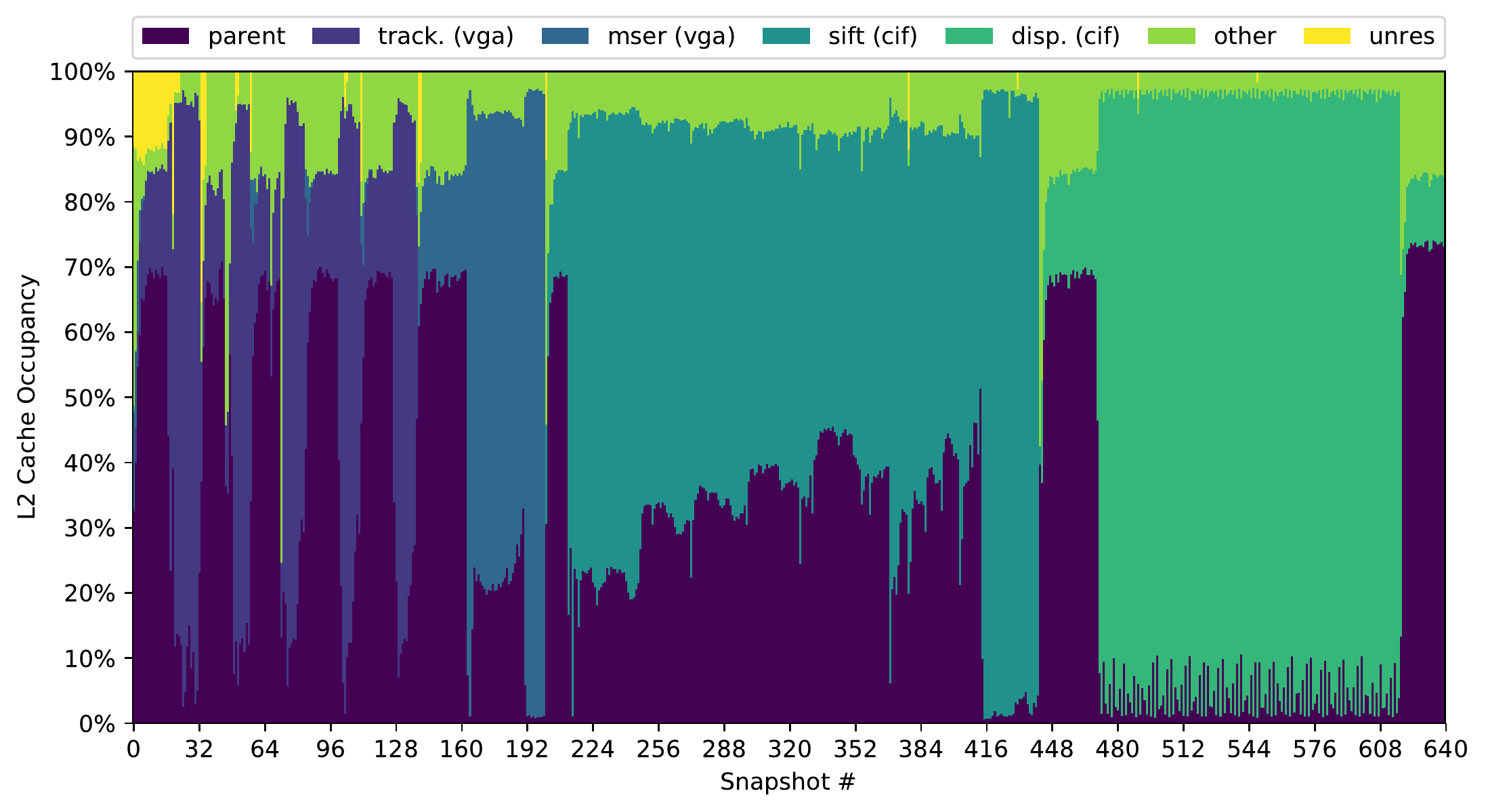}
  \vspace{-0.7cm}
  \caption{Single-core execution of 4 SD-VBS benchmarks with
    fixed-priority real-time scheduler.}
  \label{fig:4sdvbs_rt}
  \vspace{-0.4cm}
\end{figure}

To understand the impact of scheduling policies on cache utilization,
we conduct a similar experiment where we use a fixed-priority
real-time scheduler. Specifically, we run the benchmarks with
\texttt{SCHED\_FIFO} policy and set their priorities to be in
decreasing order --- i.e. \track has highest priority, \disp has
lowest priority. With this setup, we obtain
Figure~\ref{fig:4sdvbs_rt}. In this case, it is clear that the
execution of the benchmarks is performed in strict order with no
interleaving. It can also be noted how the benchmarks under analysis
vary their WSS as they progress. Another interesting observation is
that overall the execution takes less time (640 snapshots) compared to
Figure~\ref{fig:4sdvbs_nrt}, likely due to better cache locality in
absence of frequent context switching.

\begin{figure}
  \centering
  \includegraphics[width=\linewidth]{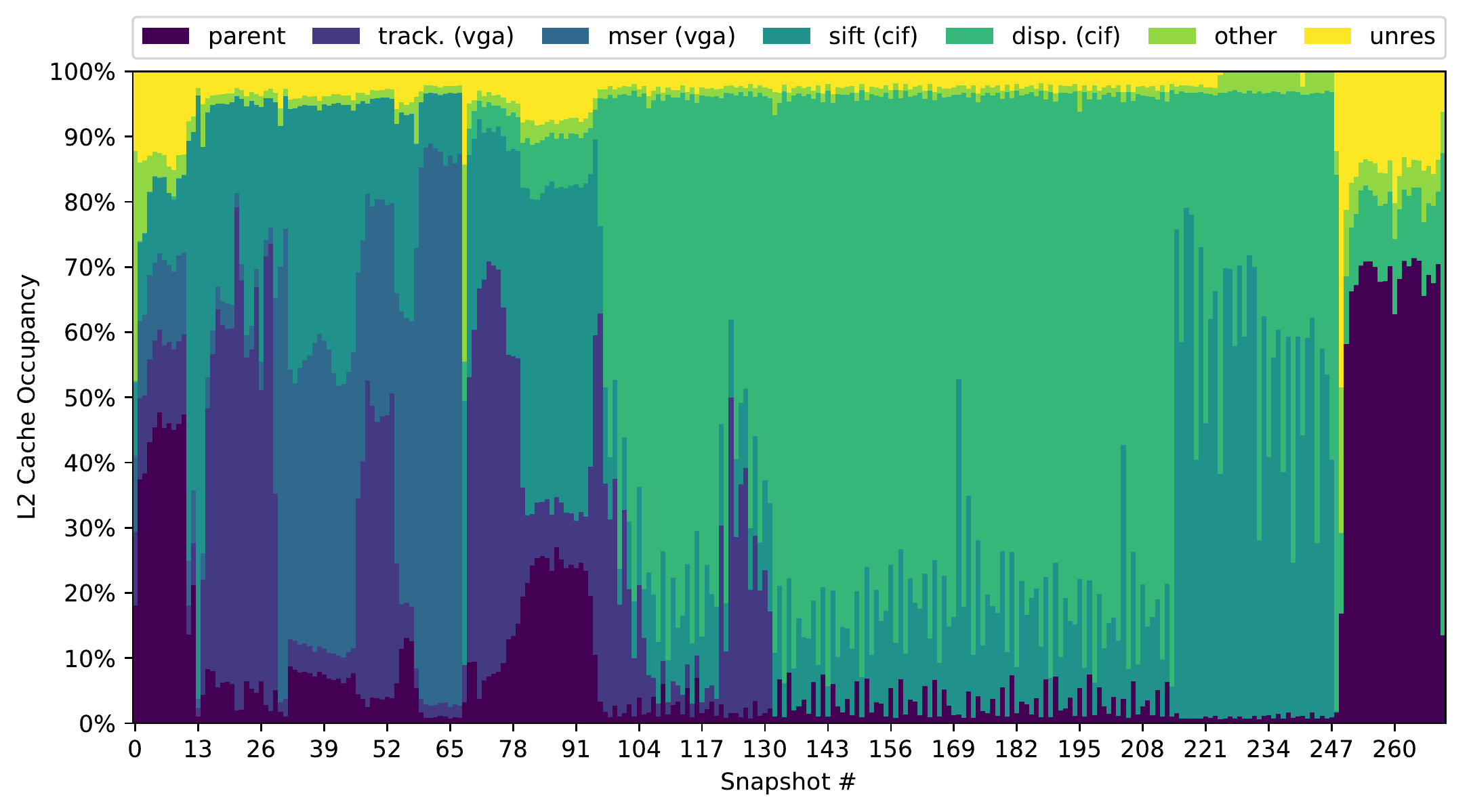}
  \vspace{-0.7cm}
  \caption{Execution of 4 SD-VBS benchmarks with fixed-priority
    real-time scheduler on 3 cores.}
  \label{fig:3c_4sdvbs}
  \vspace{-0.4cm}
\end{figure}

Next, we demonstrate that \cachedump can also be used to investigate
the behavior of the shared L2 cache even with multiple cores being
active. In this setup, we turn on 3 out of 4 cores\footnote{We also
  conducted the experiment on the full 4-core setup, which yielded
  similar results and is omitted due to space constraints.} and deploy
the same set of 4 SD-VBS benchmarks with the arrangement of real-time
priorities. We do not control process-to-CPU assignment and let the
Linux scheduler allocate processes to cores at runtime. What emerges
in terms of per-process L2 occupancy is depicted in
Figure~\ref{fig:3c_4sdvbs}. Given that L2 is a shared cache, we see
multiple applications competing for cache space. As they progress, the
amount of occupied cache depends not only on their current WSS, but
also on the WSS of co-running applications. This is particularly clear
when looking at the interplay in cache between the \track and \mser
benchmarks compared to what we observed in
Figure~\ref{fig:4sdvbs_rt}. Because only 3 cores are available, it can
also be noted that \disp only starts executing at snapshot 68, once
again significantly dominating cache utilization in the central
portion of its execution (snap.~98-215).

\subsection{\rev{Can Clash on Shared Cache be Predicted?}}
\label{sec:ev_compat}

\rev{In this section, we investigate if cache snapshotting can be used
  to predict clash on cache resources between co-running
  applications. To conduct this analysis, we first analyze the
  behavior of our applications in isolation. We use snapshotting to
  derive two type of profiles on the SD-VBS applications under
  analysis. Both profiles are constructed from snapshots acquired in
  flush mode. In the first profile, we study the sheer size of data
  allocated in cache at each snapshot --- ``Active'' set analysis. In
  the second type of profile, namely ``Reused'' set analysis, we
  evaluate only the number of lines that are the same and present in
  two successive snapshots. The reused set captures the amount of data
  that, if evicted, causes a penalty in execution time. While active
  set analysis could be carried out by carefully interpreting cache
  performance counters, reused set analysis can only be conducted by
  leveraging the exact knowledge of what lines are cached from time to
  time. Hence, this type of analysis was previously limited to
  simulation studies.}

\begin{figure}
  \centering
  \includegraphics[width=\linewidth]{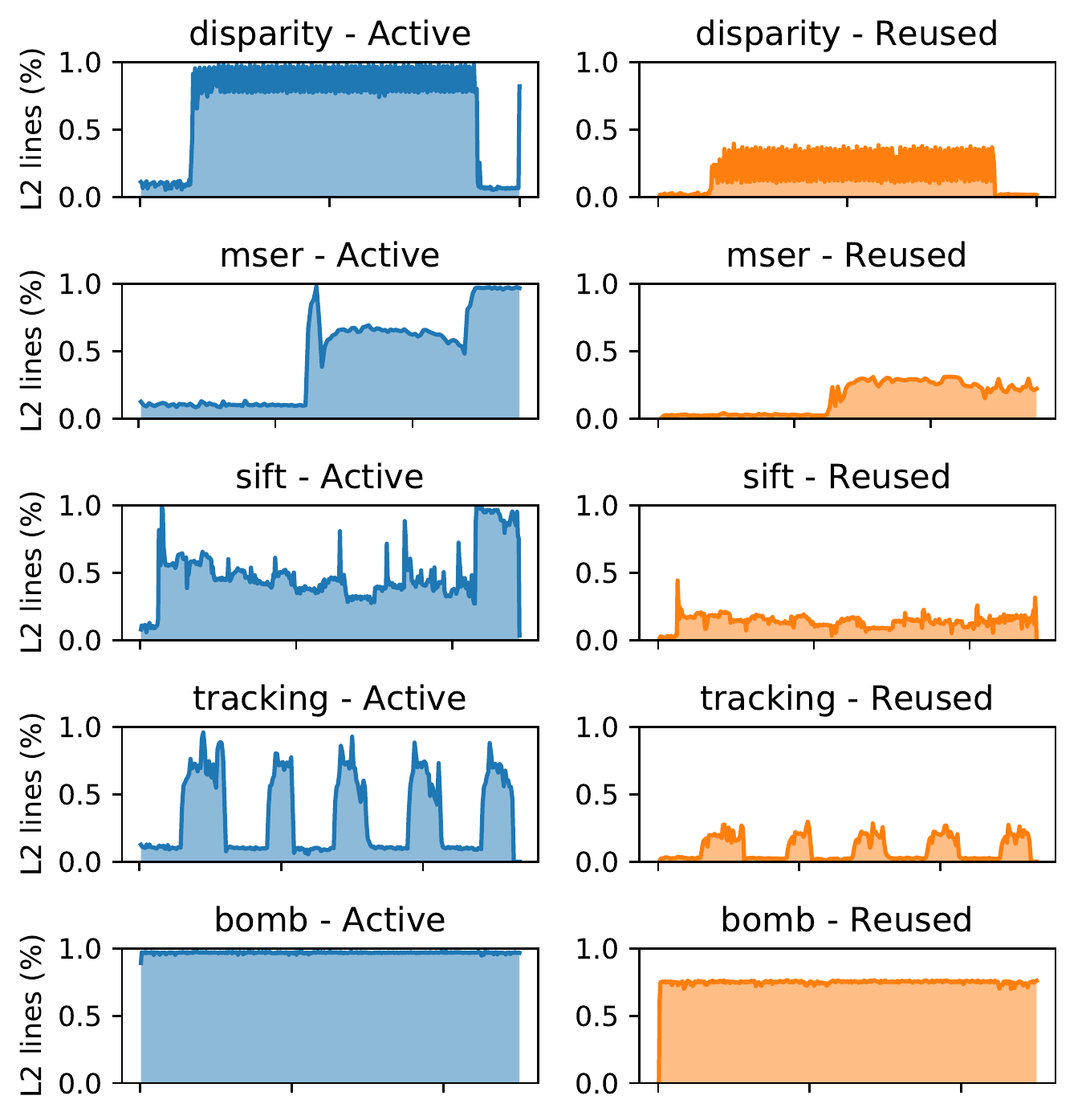}
  \vspace{-0.7cm}
  \caption{\rev{Active set (left) and reused set (right) analysis of
      SD-VBS and \bomb benchmarks.}}
  \label{fig:sdvbs_charact}
  \vspace{-0.4cm}
\end{figure}

\rev{Figure~\ref{fig:sdvbs_charact} depicts the two profiles for each
  of the considered real benchmarks. In addition, we considered a
  synthetic benchmark called \bomb which continually and sequentially
  accesses a 2.5~MB buffer. From the figure, it appears that while
  there exists a positive correlation between the active and reuse
  set, the two often (and substantially) differ.}

\rev{Using the profiles, we build two metrics to try and predict the
  impact that two applications running in parallel would have on each
  other. For this experiment, we establish the ground truth by
  observing the slowdown suffered by an application under analysis
  when running in parallel with an interfering application. In these
  measurements, \cachedump is not used. The results are reported in
  Table~\ref{tab:interf_matrix}, in the first group of rows.}

\rev{The first metric, namely ``Active Set Excess'' is based on active
  sets only. In this case, we consider two applications at a time: an
  observed, and an interfering one. We then consider the sum of their
  active sets on a per-snapshot basis, and compute how much that sum
  exceeds the size of the cache (e.g. by 0\%, by 50\% etc.). We then
  average this quantity over the full length of the profile. The
  results obtained from applying this metric on all the considered
  SD-VBS applications are reported in Table~\ref{tab:interf_matrix}
  --- second group of rows.}

\rev{A second metric, namely ``Reused Set Eviction'', considers how
  much of the reused set of the application under analysis is
  potentially evicted by an interfering application. To build this
  metric, we once again reason on a per-snapshot basis. We multiply
  the reused set quota of the application under analysis by the active
  set quota of the interfering application. The average over the full
  length of the profile is then computed. The third group of rows in
  Table~\ref{tab:interf_matrix} reports the reused set eviction metric
  computed for the considered benchmarks.}

\rev{Finally, we computed the correlation between the two metrics
  described above and the ground truth on the measured slowdown. The
  reused set eviction metric revealed an 80\% correlation with
  slowdown. It outperformed the active set excess metric which
  achieved 74\% correlation. These results serve as a proof-of-concept
  that \cachedump can be used, for instance, to perform
  interference-aware scheduling decisions.}


\begin{table}[]
  \caption{Correlation of Slowdown and Cache Activity-Based Indices}
  \begin{tabular}{cl|cccc}
    \multicolumn{1}{l}{}                                                                 &                                                                             & \multicolumn{4}{c}{\textsc{Benchmark Under Analysis}}                                       \\ \cline{3-6}
    \multicolumn{1}{l}{}                                                                 & \multirow{-2}{*}{\begin{tabular}[c]{@{}l@{}}\textsc{Interf.}\\ \textsc{Applic.}\end{tabular}} & \cellcolor[HTML]{FFCE93}{\bf \disp} & {\bf \mser} & \cellcolor[HTML]{FFCE93}{\bf \sift} & {\bf \track} \\ \hline
    & \disp                                                                   & \cellcolor[HTML]{FFCE93}1.16      & 1.13 & \cellcolor[HTML]{FFCE93}1.02 & 1.02     \\ \cline{2-6}
    & \mser                                                                        & \cellcolor[HTML]{FFCE93}1.02      & 1.06 & \cellcolor[HTML]{FFCE93}1.01 & 1.02     \\ \cline{2-6}
    & \sift                                                                        & \cellcolor[HTML]{FFCE93}1.03      & 1.05 & \cellcolor[HTML]{FFCE93}1.05 & 1.06     \\ \cline{2-6}
    & \track                                                                    & \cellcolor[HTML]{FFCE93}1.01      & 1.02 & \cellcolor[HTML]{FFCE93}1.03 & 1.06     \\ \cline{2-6}
    \multirow{-5}{*}{\begin{tabular}[c]{@{}c@{}}Slowdown\\ ($\times$)\end{tabular}}             & \bomb                                                                        & \cellcolor[HTML]{FFCE93}1.12      & 1.20 & \cellcolor[HTML]{FFCE93}1.05 & 1.04     \\ \hline \hline
    & \disp                                                                   & \cellcolor[HTML]{FFCE93}0.55      & 0.33 & \cellcolor[HTML]{FFCE93}0.19 & 0.19     \\ \cline{2-6}
    & \mser                                                                        & \cellcolor[HTML]{FFCE93}0.12      & 0.25 & \cellcolor[HTML]{FFCE93}0.04 & 0.03     \\ \cline{2-6}
    & \sift                                                                        & \cellcolor[HTML]{FFCE93}0.24      & 0.14 & \cellcolor[HTML]{FFCE93}0.14 & 0.07     \\ \cline{2-6}
    & \track                                                                    & \cellcolor[HTML]{FFCE93}0.12      & 0.05 & \cellcolor[HTML]{FFCE93}0.04 & 0.15     \\ \cline{2-6}
    & \bomb                                                                        & \cellcolor[HTML]{FFCE93}0.64      & 0.42 & \cellcolor[HTML]{FFCE93}0.46 & 0.31     \\ \cline{2-6}
    \multirow{-6}{*}{\begin{tabular}[c]{@{}c@{}}Active Set \\ Excess (\%)\end{tabular}}  & \multicolumn{4}{r}{\textsc{Correlation:}}  & \textbf{0.74}  \\ \hline
    & \disp                                                                   & \cellcolor[HTML]{FFCE93}0.15      & 0.12 & \cellcolor[HTML]{FFCE93}0.08 & 0.07     \\ \cline{2-6}
    & \mser                                                                        & \cellcolor[HTML]{FFCE93}0.03      & 0.10 & \cellcolor[HTML]{FFCE93}0.02 & 0.01     \\ \cline{2-6}
    & \sift                                                                        & \cellcolor[HTML]{FFCE93}0.08      & 0.08 & \cellcolor[HTML]{FFCE93}0.07 & 0.04     \\ \cline{2-6}
    & \track                                                                    & \cellcolor[HTML]{FFCE93}0.04      & 0.04 & \cellcolor[HTML]{FFCE93}0.03 & 0.06     \\ \cline{2-6}
    & \bomb                                                                        & \cellcolor[HTML]{FFCE93}0.18      & 0.15 & \cellcolor[HTML]{FFCE93}0.13 & 0.09     \\ \cline{2-6}
    \multirow{-6}{*}{\begin{tabular}[c]{@{}c@{}}Reused Set \\ Eviction (\%)\end{tabular}}  & \multicolumn{4}{r}{\textsc{Correlation:}}  & \textbf{0.80}  \\ \hline
      \end{tabular}
      \label{tab:interf_matrix}
      \vspace{-0.5cm}
      \end{table}

\subsection{\rev{Can we Study the Cache Replacement Policy?}}
\label{sec:ev_repl}

\rev{The last aspect we evaluated was the capability introduced by
  \cachedump to evaluate the replacement policy implemented by the
  hardware. We focus on two aspects. First, we validate that the
  policy indeed follows random replacement. Second, we study how well
  the implemented replacement policy matches a truly random
  replacement. To conduct these experiments we devised a special
  synthetic benchmark, namely \repl. The \repl benchmark allocates
  from user-space a set of contiguous physical pages with a total size
  equal to the L2 cache size. This is done by \texttt{mmap} leveraging
  support for huge pages --- \texttt{MAP\_HUGE\_2MB} flag.}

\begin{figure}
  \centering
  \includegraphics[width=\linewidth]{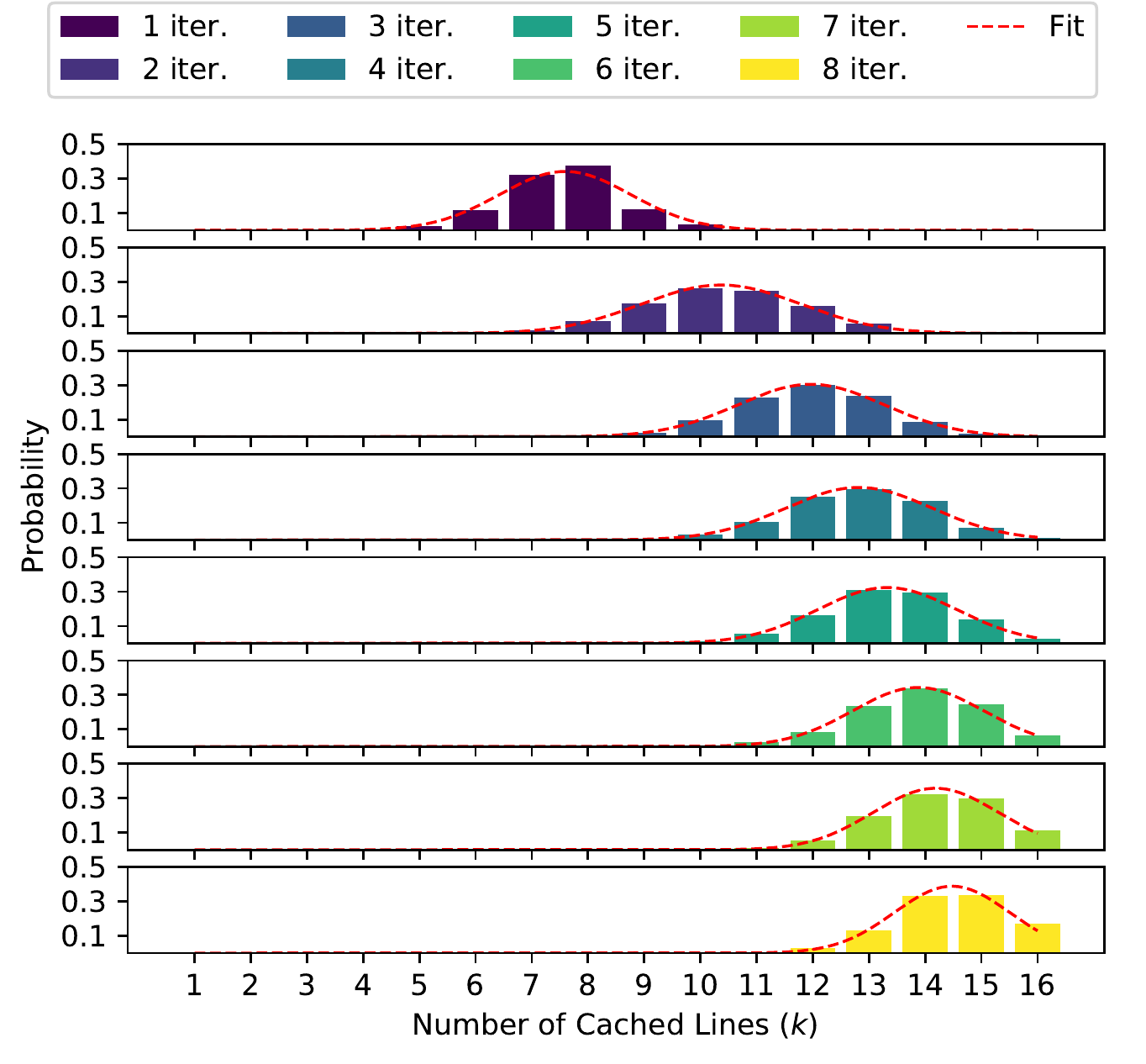}
  \vspace{-0.7cm}
  \caption{\rev{Probability of $k$ lines on the same cache set being
      cached after having been accessed 1 (top) through 8 (bottom)
      times.}}
  \label{fig:repl} 
  \vspace{-0.4cm}
\end{figure}

\rev{Because the allocated buffer is aligned with the cache size, the
  first line, say Line~$A_1$, of the buffer necessarily maps to cache
  set~0. Similarly, the line (Line~$A_2$) that is exactly 128~KB away
  (the size of one cache way) from Line~$A_1$ also maps to set 0. With
  a similar reasoning we identify a set of 16 lines $\{A_1, \ldots,
  A_{16}\}$ that map to set 0. If the cache implements a deterministic
  cache replacement policy (e.g. LRU or FIFO), then accessing lines
  $A_1$ through $A_{16}$ will result in a snapshot containing all the
  lines. Following this idea, the \repl benchmark touches lines $A_1$
  through $A_{16}$ a configurable number of times (iterations). Then,
  it delivers a signal to the \trigger (event-based activation) to
  acquire a snapshot in transparent mode. In the snapshot, we evaluate
  how many of the 16 lines are actually present in cache. By repeating
  the same experiment 1000 times, we can plot the probability density
  that $k \in \{1, \ldots, 16\}$ out of 16 lines are found in
  cache. We repeat the experiment by performing from 1 to 8 iterations
  over lines $A_1$ through $A_{16}$ before acquiring a snapshot. The
  resulting density plots are reported in Figure~\ref{fig:repl}.}

\begin{figure}
  \centering \includegraphics[width=\linewidth]{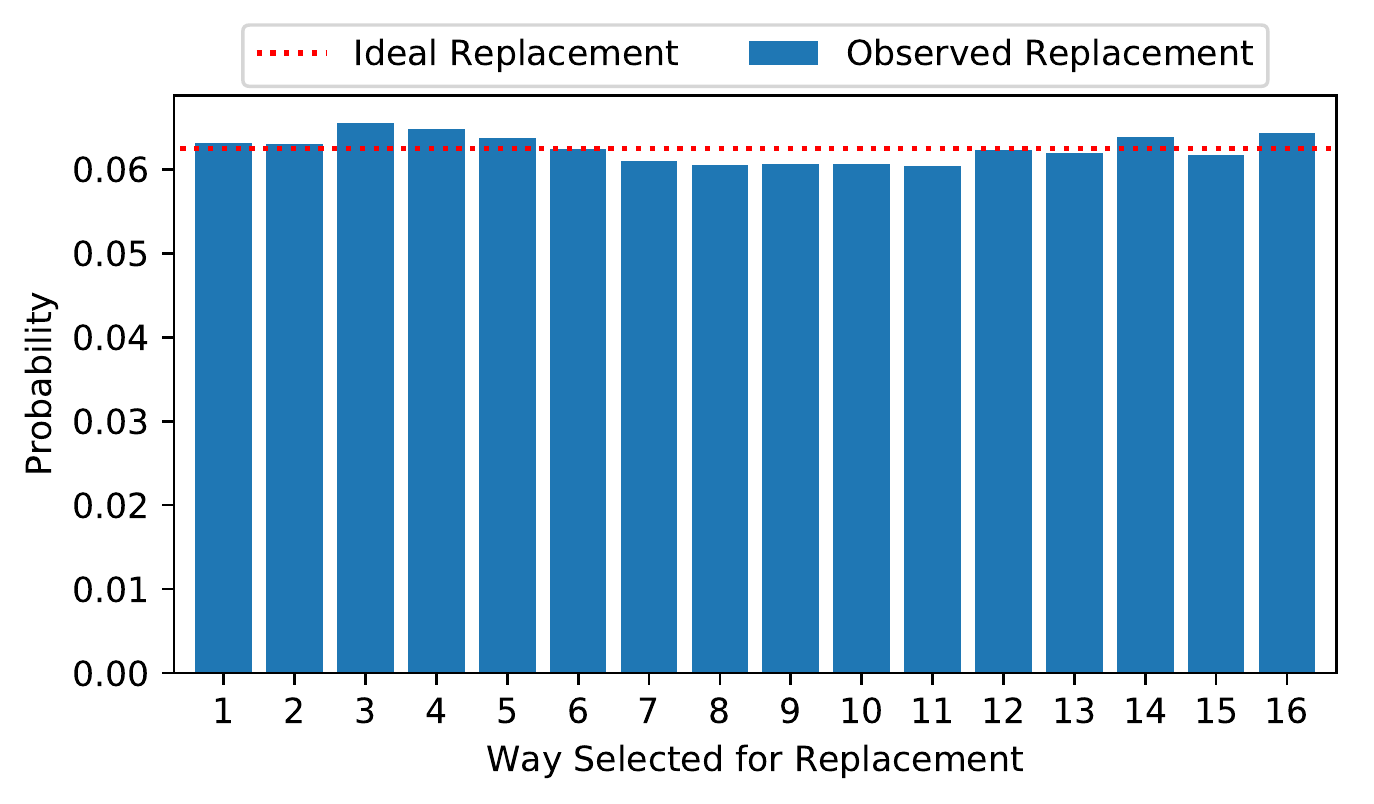}
  \vspace{-0.7cm}
  \caption{\rev{Probability of cache way 1-16 being selected for
      replacement.}}
  \label{fig:victim}
  \vspace{-0.4cm}
\end{figure}

\rev{In our last experiment, we evaluate how closely the implemented
  random replacement policy matches a truly random replacement policy.
  To evaluate this aspect, we use a variant of the \repl benchmark. We
  consider again lines $A_1$ through $A_{16}$. But in this case, the
  benchmark activates the \trigger after touching each line and stops
  after reaching line $A_{16}$. A single run produces 16
  snapshots. Moreover in snapshot 1, we can derive which cache way was
  selected to allocate $A_1$, in snapshot 2 what way was selected for
  $A_2$ and so on. We run the experiment 2000 times and collect a
  total of 32,000 replacement decisions. We then compute the number of
  times each of the 16 cache ways were selected for allocation over the
  total number of observations. The results are reported in
  Figure~\ref{fig:victim}. In a perfect random replacement scheme,
  each way has probability $\frac{1}{16} = 6.25\%$ of being
  selected. The implemented replacement policy does not deviate
  significantly from a perfect replacement, although interestingly, it
  appears that the central ways (ways 7 to 11)
  are statistically less likely to be selected for allocation.}

\section{Conclusion and Future Work}
In this work, we have proposed a technique, namely \cachedump, that
leverages existing yet untapped micro-architectural support to enable
cache content snapshotting. The proposed implementation has
highlighted that \cachedump can provide unprecedented insights on
application-level features in the usage of cache resources, and on
system-level properties. As such, we envision that \cachedump and its
analogues will serve as a powerful instrument for system designers to
better understand the interplay between applications, system
components, and cache hierarchy. Ultimately, we expect that it will
complement existing simulation-based and static analysis approaches by
providing a way to refine and validate cache memory models. While we
have restricted ourselves to analyzing the LLC in this work,
\ramindex's capabilities far exceed that. Hence, we encourage the
community to extend and refine the proposed open-source implementation
to conduct a wider range of studies, e.g. on the behavior of private
caches, TLBs, and coherence controllers.

\bibliographystyle{IEEEtran}
\bibliography{biblio}


\end{document}